\documentclass[aps,showpacs,showkeys,preprint,superscriptaddress]{revtex4-1}

\usepackage[utf8]{inputenc}

\usepackage{amssymb, amsmath, units}
\usepackage[english]{babel}
\usepackage{bm}
\usepackage{graphicx}
\usepackage{color}
\usepackage{gensymb}

\usepackage{paralist}

\usepackage[titletoc]{appendix}

\usepackage{hyperref}

\usepackage[table]{xcolor} 
\usepackage{booktabs}
\usepackage{subfig}
\allowdisplaybreaks

\DeclareMathAlphabet{\mathsfit}{\encodingdefault}{\sfdefault}{m}{sl}
\DeclareMathAlphabet{\mathsfitbf}{\encodingdefault}{\sfdefault}{bx}{sl}

\newcommand{\bra}[1]{\langle #1 \rvert}
\newcommand{\ket}[1]{\lvert #1 \rangle}

\DeclareMathOperator{\trace}{tr}

\DeclareMathOperator{\im}{Im}

\newcommand{\e}[0]{\text{e}}

\renewcommand{\vec}[1]{\bm{#1}}
\newcommand{\V}[4]{V{}^{#1}_{#3}{\;}^{#2}_{#4}}
\newcommand{\mathcalV}[4]{\mathcal{V}{}^{#1}_{#3}{\;}^{#2}_{#4}}
\newcommand{\vb}[0]{\text{v}}
\newcommand{\cb}[0]{\text{c}}
\newcommand{\Ho}[0]{\text{H}}
\newcommand{\Lu}[0]{\text{L}}

\begin{document}

\author{Judith Specht}
\email[]{specht@itp.tu-berlin.de}
\affiliation{Institut f\"ur Theoretische Physik, Nichtlineare Optik und Quantenelektronik, Technische Universit\"at Berlin, Hardenbergstr. 36, 10623 Berlin, Germany}

\author{Eike Verdenhalven}
\affiliation{Institut f\"ur Theoretische Physik, Nichtlineare Optik und Quantenelektronik, Technische Universit\"at Berlin, Hardenbergstr. 36, 10623 Berlin, Germany}

\author{Bj\"orn Bieniek}
\affiliation{Fritz-Haber-Institut der Max-Planck-Gesellschaft, Berlin, Germany}

\author{Patrick Rinke}
\affiliation{Fritz-Haber-Institut der Max-Planck-Gesellschaft, Berlin, Germany}
\affiliation{COMP Department of Applied Physics, Aalto University, P.O. Box 11100, Aalto FI-00076, Finland}

\author{Andreas Knorr}
\affiliation{Institut f\"ur Theoretische Physik, Nichtlineare Optik und Quantenelektronik, Technische Universit\"at Berlin, Hardenbergstr. 36, 10623 Berlin, Germany}

\author{Marten Richter}
\email[]{marten.richter@tu-berlin.de}
\affiliation{Institut f\"ur Theoretische Physik, Nichtlineare Optik und Quantenelektronik, Technische Universit\"at Berlin, Hardenbergstr. 36, 10623 Berlin, Germany}

\title{Theory of excitation transfer between two-dimensional semiconductor and molecular layers}

\begin{abstract}
The geometry-dependent energy transfer rate from an electrically pumped inorganic semiconductor quantum well into an organic molecular layer is studied theoretically. We focus on F\"orster-type nonradiative excitation transfer between the organic and inorganic layer and include quasi-momentum conservation and intermolecular coupling between the molecules in the organic film. (Transition) partial charges calculated from density-functional theory are used to calculate the coupling elements. The partial charges describe the spatial charge distribution and go beyond the common dipole-dipole interaction. We find that the transfer rates are highly sensitive to variations in the geometry of the hybrid inorganic/organic system. For instance, the transfer efficiency is improved by orders of magnitude by tuning the relative orientation and positioning of the molecules. Also, the operating regime is identified where in-scattering dominates over unwanted back-scattering from the molecular layer into the substrate.
\end{abstract}

\date{\today}
\maketitle

\section{Introduction}

A potential advantage of hybrid inorganic/organic systems over their individual constituents is that a synergistic combination can lead to novel optoelectronic properties and tunable functionality \cite{Basko:EurPhysJB:99, Blumstengel:PhysRevLett:06, Itskos:PhysRevB:07, Neves:AdvancedFunctionalMaterials:08, Liang:ThinSolidFilms:13, Schlesinger:NatCommun:15, Qiao:ACSNano:15, Friede:PhysRevB:15, Ljungberg:NewJPhys:17}. Typical components include organic materials such as organic dye molecules and inorganic semiconductor nanostructures such as a quantum well (QW) or a semiconductor surface \cite{Haug::04, Nakamura:JapanJApplPhys:95, Zimmermann:PureApplChem:97, Zimmermann::03, Singh:PhysRevB:17}. Inorganic semiconductors have several favorable properties such as high charge carrier mobilities and efficient carrier injection that could be benefitially combined with the strong light-matter coupling high radiative emission efficiency of organic molecules.  Furthermore, novel types of excitation processes can emerge in hybrid nanostructures, e.g., Frenkel-Wannier excitons \cite{Agranovich:JPhys:CondensMatter:98, Richter:PhysRevMaterials:17} and hybrid charge transfer interface states (i.e., excitons with the electron and hole located at different constituents of the inorganic/organic heterostructure) \cite{Vaynzof:PhysRevLett:12, Piersimoni:JPhysChemLett:15}. In this work, we study dipole-induced excitation transfer pathways from a strongly electrically pumped inorganic substrate across the two-dimensional interface towards the organic layer on a microscopic level.

F\"orster-type nonradiative energy transfer \cite{Foerster:AnnalenderPhysik:48} is such a process. It can couple electronic states in an inorganic semiconductor nanostructure to Frenkel excitons in the organic component and dominates if wave function overlap between the organic and inorganic layer is negligible. Such excitation transfer processes have been the object of experimental \cite{Blumstengel:PhysRevLett:06, Blumstengel:NewJPhys:08, Itskos:PhysRevB:07, Plehn:JPhysChemB:15, Qiao:ACSNano:15} as well as theoretical studies \cite{Agranovich:SolidStateCommun:94, Agranovich:JPhys:CondensMatter:98, Verdenhalven:PhysRevB:14}. However, since a thorough experimental characterization of the underlying microscopic coupling mechanisms is difficult, theoretical studies can extend knowledge towards a detailed understanding of the excitation transfer dynamics in hybrid systems.

In this work, we use a density matrix formalism (similar to Ref.~\onlinecite{Verdenhalven:PhysRevB:14}) to study the excitation transfer in the composite inorganic-organic system. For the interactions, we include interlayer Coulomb coupling (F\"orster-type and electrostatic) as well as intermolecular coupling within the organic layer. The intermolecular coupling leads to the formation of bands in the organic system. This is particularly relevant in the case of densely packed molecular films and thus for small intermolecular distances. Besides the dipole-dipole excitation transfer contributions to the Coulomb Hamiltonian, all electrostatic monopole-monopole coupling terms are also considered, resulting in electrostatic shifts in the resonance energies that strongly depend on the molecular coverage. The microscopic Coulomb coupling elements are calculated using (transition) partial charges \cite{Madjet:JPhysChemB:06,Campana:JChemTheoryComput:09} obtained from density-functional theory (DFT). This allows a more accurate modeling of Coulomb interaction processes for small distances between the interacting constituents compared to a simple point-dipole approximation. We assume a periodic arrangement of the molecules in the organic layer as in Ref.~\onlinecite{Verdenhalven:PhysRevB:14}. In this way, it is possible to consistently treat both the semiconductor and molecular quantities using a quasi-momentum representation \cite{Verdenhalven:PhysRevB:14}.

In this paper, we derive a Coulomb-interaction excitation transfer rate from the semiconductor substrate into the molecular layer and vice versa. Our approach represents the case of exciting the optically active organic molecular layer by strong electrical pumping of the semiconductor QW. We focus on the coupling of electrically pumped, i.e. occupied electron-hole continuum states of the semiconductor to the molecular excitons. The semiconductor continuum covers a broad energy range, which makes the coupling efficiency less sensitive to resonance energy mismatches between the organic and inorganic component.
We find that the effect of interlayer coupling is determined by microscopic quasi-momentum selection rules that depend on the geometry of the hybrid structure. Therefore, the parameter studies presented in this manuscript can help to increase the energy transfer efficiency by geometry optimization of the hybrid structure.

This work is organized as follows: First, we illustrate the theoretical treatment of the model system employing a partial charge approximation of the Coulomb coupling elements (Sec.~\ref{sec:model-system}). In Sec.~\ref{sec:EOM}, equations of motion for the excitation transfer are derived for the system density operator and the scattering rate is deduced. Finally, we present the numerical evaluation using microscopic input parameters obtained from DFT calculations (Sec.~\ref{sec:TransferRates}).

\section{Model system and Hamiltonian} \label{sec:model-system}

\begin{figure}
	\hfill
	\begin{minipage}[b]{0.225\linewidth}
		\centering
		\includegraphics[width=\linewidth]{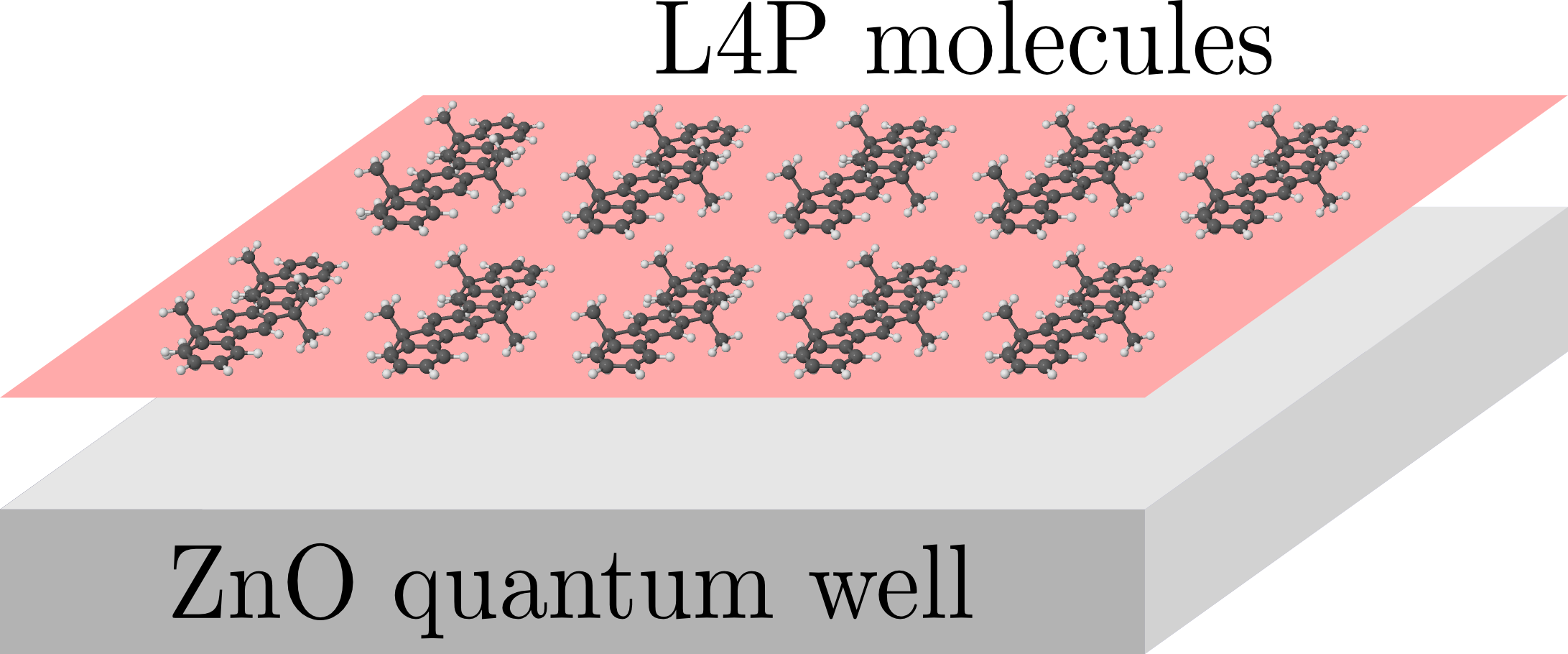}
		(a)
	\end{minipage}
	\hfill
	\begin{minipage}[b]{0.24\linewidth}
		\centering
		\includegraphics[width=\linewidth]{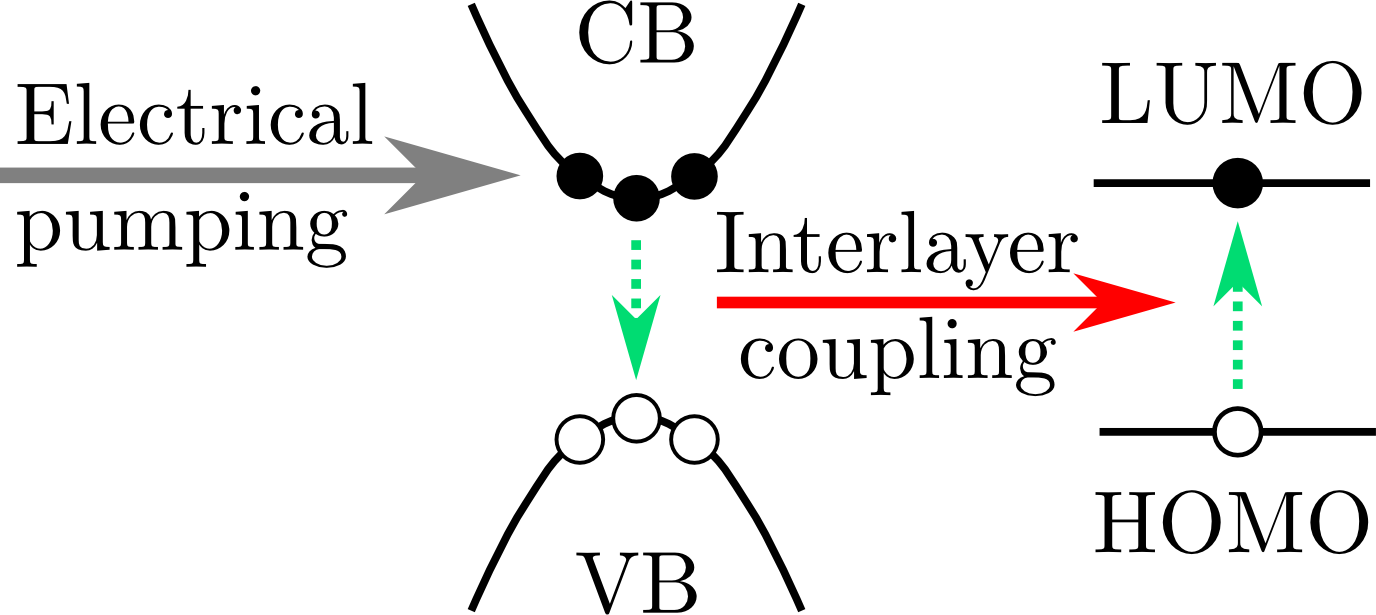}
		(b)
	\end{minipage}
	\hfill\hfill
	\caption{(a)~Model system consisting of a single layer of L4P molecules adsorbed on a ZnO substrate. (b)~Schematic of F\"orster energy transfer from an electrically pumped semiconductor to a molecule.}
	\label{fig:ModelSystem}
\end{figure}

\subsection{Model system and input parameters}

In the considered model system, flat ladder-type quarterphenyl (L4P) molecules \cite{Kobin:JMaterChem:12} are arranged in a quasi two-dimensional film on a ZnO quantum well (QW), cf. Fig.~\ref{fig:ModelSystem}(a). The microscopic input parameters are taken from  DFT calculations. The adsorption geometry is calculated for an organic film of weakly bonded L4P molecules relaxed on top of a ZnO($10\bar{1}0$) surface. This geometry is kept fixed for deriving the partial charges of the organic and inorganic component in separate DFT calculations in vacuum. The DFT calculations employ the hybrid exchange-correlation (xc)  functional HSE06 \cite{Heyd:JChemPhys:03} as implemented in the FHI-aims code \cite{Blum:ComputPhysCommun:09,Xinguo/implem_full_author_list,Levchenko2015}. We use \textit{tight} numerically settings for all calculations. Relativistic effects were accounted for by the \textit{atomic ZORA} approach \cite{Blum:ComputPhysCommun:09}. The substrate is modeled using the slab-approach, where a ZnO unit cell with a $z$ extension corresponding to a QW thickness of $\unit[4.3]{nm}$ is defined and repeated periodically. The periodic images of the slab are separated by a vacuum layer $>$30$\mathrm{\AA}$ in $z$ direction. Potential electrostatic interactions between periodic images are compensated by a dipole correction \cite{Bengtsson:PhysRevB:99}. For the integrations over the Brillouin zone for obtaining the transition partial charges we used a k-point grid with a density of pt$\cdot\mathrm{\AA}$ in the directions corresponding to the surface plane. Van der Waals interactions were taken into account by means of the Tkatchenko-Scheffler (TS) scheme \cite{Tkatchenko:PhysRevLett:09}. For the ZnO surface the TS parameterization of Ref.~\onlinecite{Zhang:PhysRevLett:11} was used (see Ref.~\onlinecite{Hofmann/etal:2013,Bieniek/Hofmann/Rinke:2015} for details).

\subsection{Hamiltonian}

We focus on the energetically lowest allowed electronic transitions between the highest occupied molecular orbital (HOMO, $\Ho$) and lowest unoccupied molecular orbital (LUMO, $\Lu$) in the molecules and between the valence and conduction band in the semiconductor substrate. The Hamilton operator of the hybrid system for calculating the transfer rate consists of three parts:
$\hat{H} = \hat{H}_0 + \hat{H}_{\text{C}}^{\text{m-m}} + \hat{H}_{\text{C}}^{\text{m-s}}$.

The Hamiltonian
\begin{equation} \label{eq:FreePart}
	\hat{H}_0 = \hat{H}_0^{\text{m}} + \hat{H}_0^{\text{s}} = \sum_{A, \nu} \varepsilon_{A}^{\nu} \, \hat{a}_{A, \nu}^{\dagger} \hat{a}_{A, \nu}^{\phantom{\dagger)}} +
	\sum_{\lambda, \vec{k}} \varepsilon_{\lambda}^{\vec{k}} \, \hat{a}_{\lambda, \vec{k}}^{\dagger} \hat{a}_{\lambda, \vec{k}}^{\phantom{\dagger}}
\end{equation}
contains the free-particle energies $\varepsilon_{A}^{\nu}$ and $\varepsilon_{\lambda}^{\vec{k}}$ of the carriers in the molecular layer and in the semiconductor bands, respectively. The index $A$ (running over $\Ho$ and $\Lu$) denotes the molecular orbital of the $\nu$-th molecule with electronic wave function $\psi_{A, \nu}$. Since we assume identical molecules, $\varepsilon_{A}^{\nu} \equiv \varepsilon_{A}$ holds. $\lambda$ includes the valence ($\vb$) and conduction ($\cb$) band and $\vec{k} \equiv \vec{k}_{\parallel}$ the (two-dimensional) wave vector of a semiconductor electron in the two-dimensional QW plane with wave function $\psi_{\lambda, \vec{k}}$ in envelope-function approximation \cite{Haug::04}.
$\hat{a}_{A, \nu}^{(\dagger)}$ and $\hat{a}_{\lambda, \vec{k}}^{(\dagger)}$ are the annihilation (creation) operators for an electron in a molecule and in the semiconductor QW, respectively.

We consider two different contributions from the Coulomb interaction: First, $\hat{H}_{\text{C}}^{\text{m-m}}$ describes the intermolecular coupling between molecules in the organic layer:
\begin{align}
	\begin{split} \label{eq:m-mCoulombHamiltonianPositionSpace}
		\hat{H}_{\text{C}}^{\text{m-m}} = & \frac{1}{2} \sum_{A, B} \sum_{\nu_a \neq \nu_b} \V{A, \nu_a}{B, \nu_b}{A, \nu_a}{B, \nu_b} \,
		\hat{a}_{A, \nu_a}^{\dagger} \hat{a}_{B, \nu_b}^{\dagger} \hat{a}_{B, \nu_b}^{\phantom{\dagger}} \hat{a}_{A, \nu_a}^{\phantom{\dagger}} \\
		& + \sum_{\nu_a \neq \nu_b} \V{\Ho, \nu_a}{\Lu, \nu_b}{\Lu, \nu_a}{\Ho, \nu_b} \,
		\hat{a}_{\Ho, \nu_a}^{\dagger} \hat{a}_{\Lu, \nu_b}^{\dagger} \hat{a}_{\Ho, \nu_b}^{\phantom{\dagger}} \hat{a}_{\Lu, \nu_a}^{\phantom{\dagger}}
	\end{split}
\end{align}
with the Coulomb coupling matrix element
\begin{align}
	\begin{split} \label{eq:m-mCouplingElementPositionSpace}
		\V{A, \nu_a}{B, \nu_b}{A^\prime, \nu_a}{B^\prime, \nu_b} = & \int \text{d}^3r \int \text{d}^3r^\prime \; \psi_{A, \nu_a}^* (\vec{r}) \psi_{B, \nu_b}^* (\vec{r}^\prime) \\
		& \times e^2 G^\text{m-m} (\vec{r}, \vec{r}^\prime) \psi_{B^\prime, \nu_b}^{\phantom{*}} (\vec{r}^\prime) \psi_{A^\prime, \nu_a}^{\phantom{*}} (\vec{r}).
	\end{split}
\end{align}
$G^\text{m-m} (\vec{r}, \vec{r}^\prime)$ denotes the Green's functions for the Coulomb interaction between two charges at $\vec{r}$ and $\vec{r}^\prime$, as it arises from Poisson's equation for interacting charges and is discussed later in this section.
Note that we distinguish two contributions in Eq.~\eqref{eq:m-mCoulombHamiltonianPositionSpace} \cite{Richter:PhysStatusSolidiB:06}: The diagonal monopole-monopole coupling (first term) represents the electrostatic Coulomb interaction between charge densities and gives rise to an energy renormalization of the electronic states. The off-diagonal F\"orster coupling (second term) describes an excitation energy transfer.

Second, the molecule-semiconductor (interlayer) Coulomb Hamiltonian $\hat{H}_{\text{C}}^{\text{m-s}}$ describes the coupling between the molecules and the electrons in the semiconductor substrate, cf. Fig.~\ref{fig:ModelSystem}(b):
\begin{align}
	\begin{split} \label{eq:m-sCoulombHamiltonianPositionSpace}
		\hat{H}_{\text{C}}^{\text{m-s}} & = \sum_{\lambda, \vec{k}, \vec{k}^\prime} \sum_{A, \nu} \V{\lambda, \vec{k}}{A, \nu}{\lambda, \vec{k}^\prime}{A, \nu} \,
		\hat{a}_{\lambda, \vec{k}}^{\dagger} \hat{a}_{A, \nu}^{\dagger} \hat{a}_{A, \nu}^{\phantom{\dagger}} \hat{a}_{\lambda, \vec{k}^\prime}^{\phantom{\dagger}} \\
		& + \Bigl( \sum_{\vec{k}, \vec{k}^\prime} \sum_{\nu} \V{\cb, \vec{k}}{\Ho, \nu}{\vb, \vec{k}^\prime}{\Lu, \nu} \,
		\hat{a}_{\cb, \vec{k}}^{\dagger} \hat{a}_{\Ho, \nu}^{\dagger} \hat{a}_{\Lu, \nu}^{\phantom{\dagger}} \hat{a}_{\vb, \vec{k}^\prime}^{\phantom{\dagger}} + \textit{h.c.} \Bigr),
	\end{split}
\end{align}
\begin{align}
	\begin{split} \label{eq:m-sCouplingElement}
		\V{\lambda, \vec{k}}{A, \nu}{\lambda^\prime, \vec{k}^\prime}{B, \nu} & = \int \text{d}^3r \int \text{d}^3r^\prime \;
		\psi_{\lambda, \vec{k}}^* (\vec{r}) \psi_{A, \nu}^* (\vec{r}^\prime) \\
		& \times e^2 G^\text{m-s} (\vec{r}, \vec{r}^\prime) \psi_{B, \nu}^{\phantom{*}} (\vec{r}^\prime) \psi_{\lambda^\prime, \vec{k}^\prime}^{\phantom{*}} (\vec{r}).
	\end{split}
\end{align}
Couplings such as Coulomb interaction between the semiconductor electrons within the QW substrate are not considered here, since our goal is to describe strong, incoherent electrical pumping of the semiconductor with large carrier densities, which will suppress the formation of Wannier-exciton like bound states within the semiconductor.

The dielectric screening in the composite system is taken into account by introducing effective dielectric constants $\epsilon_\text{eff}^\text{m-m} = \epsilon_\text{m} (\epsilon_\text{s} + \epsilon_\text{m})/(\epsilon_\text{s} - \epsilon_\text{m})$ for the intermolecular and $\epsilon_\text{eff}^\text{m-s} = \tfrac{1}{2} (\epsilon_\text{s} + \epsilon_\text{m})$ for the interlayer Coulomb matrix elements. The effective dielectric constants are derived treating two half spaces with different bulk dielectrics, $\epsilon_\text{m}$ in the molecular layer and $\epsilon_\text{s}$ in the semiconductor substrate. Electrostatic charges within one of the half spaces will influence the electrostatic potential in the other half space, which can be described using the concept of image charges, cf. Refs.~\onlinecite{Jackson::99, Specht:ProcSPIE:16}. The Green's functions are given by
\begin{align}
	\begin{split}
		G^\text{m-m} & (\vec{r}, \vec{r}^\prime) = \frac{1}{4 \pi \epsilon_0} \Bigl( \frac{1}{\epsilon_\text{m} \lvert \vec{r} - \vec{r}^\prime \rvert} \\
		- & \frac{1}{\epsilon_\text{eff}^\text{m-m} \sqrt{(x-x^\prime)^2 + (y-y^\prime)^2 + (z+z^\prime)^2}} \Bigr)
	\end{split}
\end{align}
for the intermolecular and
\begin{equation}
	G^\text{m-s} (\vec{r}, \vec{r}^\prime) = \frac{1}{4 \pi \epsilon_0 \epsilon_\text{eff}^\text{m-s}} \frac{1}{\lvert \vec{r} - \vec{r}^\prime \rvert}
\end{equation}
for the interlayer Coulomb interaction.

\subsection{Partial charge approximation} \label{sec:PCA}

A dipole approximation is a common procedure for deriving Coulomb coupling elements for excitation energy transfer (cf. e.g. Refs.~\onlinecite{Lovett:PhysRevB:03,Danckwerts:PhysRevB:06, Curutchet:JPhysChemC:08, Machnikowski:PhysStatusSolidiB:09, Specht:PhysRevB:15}). However, the approximation is questionable if the size of the interacting wave functions is on the same order as the distance between the two constituents (i.e., in the case of inter-molecular coupling). One way to overcome this limitation is the extended dipole approximation \cite{Madjet:JPhysChemB:06}.
Here, a well-known method from the force-field community and quantum chemistry is adapted, using partial charges that are obtained numerically by fitting the electrostatic potential \cite{Momany:JPhysChem:78, Cox:JComputChem:81, Singh:JComputChem:84, Besler:JComputChem:90, Chirlian:JComputChem:87, Breneman:JComputChem:90, Bayly:JPhysChem:93,Sigfridsson:JComputChem:98, Hu:JChemTheoryComput:07}.

\begin{figure}
	\hfill
	\begin{minipage}{0.15\linewidth}
		\centering
		\includegraphics[width=\linewidth]{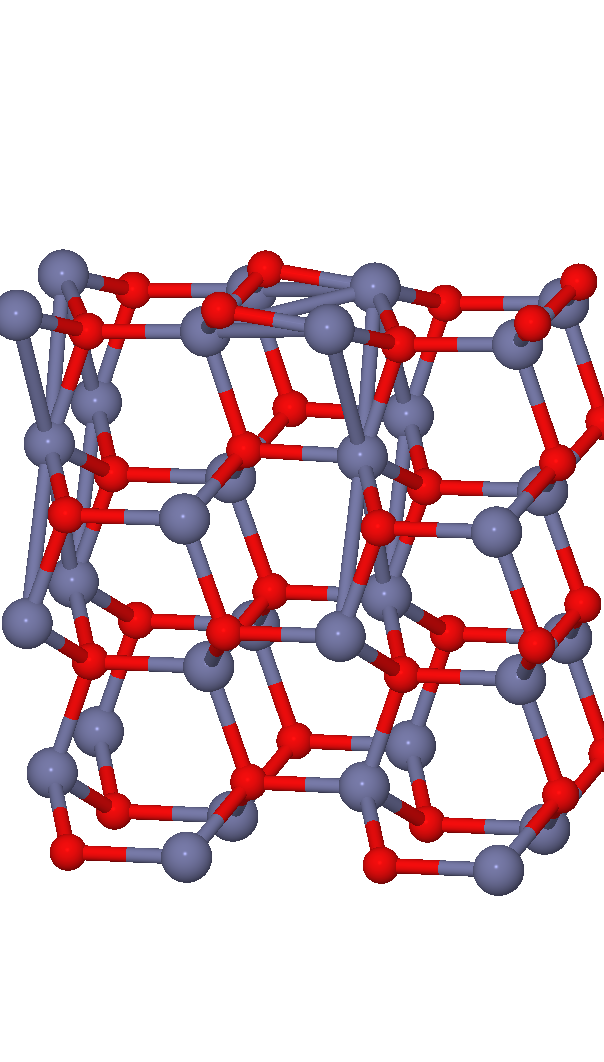}
		(a)
	\end{minipage}
	\hfill
	\begin{minipage}{0.15\linewidth}
		\centering
		\includegraphics[width=\linewidth]{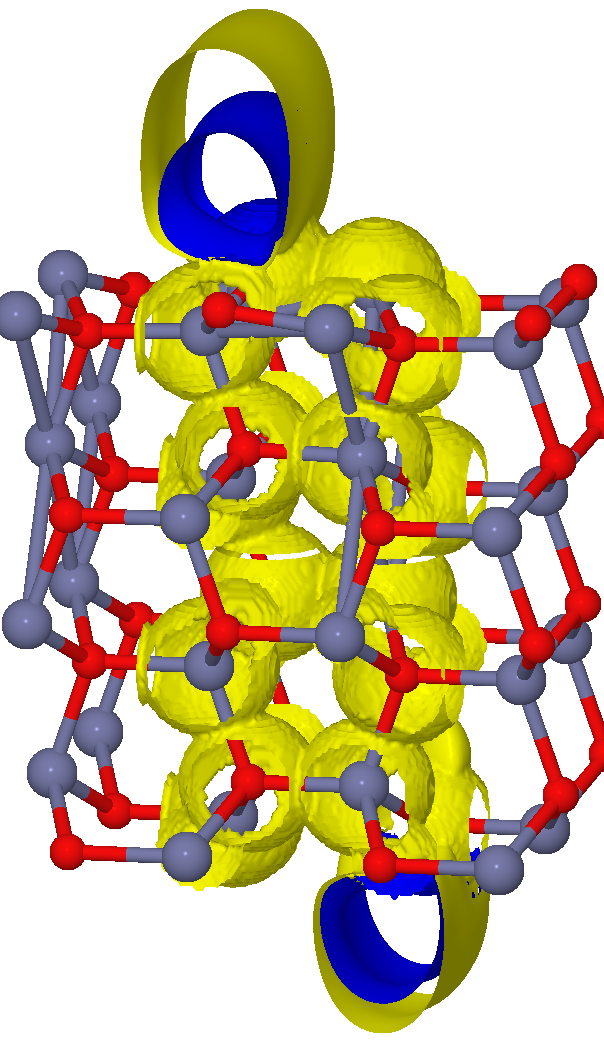}
		(b)
	\end{minipage}
	\hfill
	\begin{minipage}{0.15\linewidth}
		\centering
		\includegraphics[width=\linewidth]{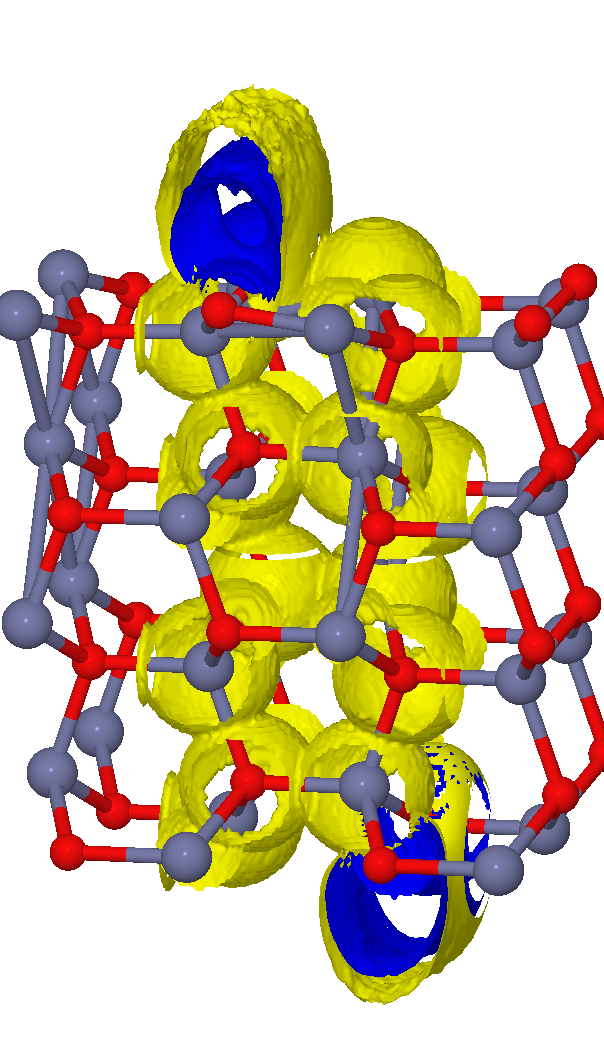}
		(c)
	\end{minipage}
	\hfill\hfill
	\caption{(a)~Atomic structure of the ZnO ($10\bar{1}0$) surface. (b)~Electrostatic potential obtained from a DFT calculation employing the hybrid xc-functional HSE06 \cite{Heyd:JChemPhys:03}. (c)~Electrostatic potential of the ZnO ($10\bar{1}0$) surface generated by the approximating partial charges. The calculated electrostatic potential of (b) is well represented by the reconstruction with partial charges. The unit cell is periodically extended perpendicular to the surface.}
	\label{fig:PotentialFit}
\end{figure}

In Ref.~\onlinecite{Madjet:JPhysChemB:06}, this partial charge technique was used to describe two strongly coupled pigments in light-harvesting complexes, each given by a many-particle wave function containing $N$ electrons. Here, we have a Hamiltonian in second quantization. We define the single-particle density $\rho_\nu^{AB} (\vec{r})$ of the $\nu$-th molecule as the product of two molecular wave functions \cite{Scholes:AnnuRevPhysChem:03}:
$ \rho_\nu^{AB} (\vec{r}) = \psi_{A, \nu}^* (\vec{r}) \psi_{B, \nu}^{\phantom{*}} (\vec{r})$.
For $A \neq B$, it represents the HOMO-LUMO transition density.
We introduce the potential solving the Poisson equation with mirror charges $\Delta_{\vec{r}}\phi_{\nu_b}^{B B^\prime} (\vec{r}) = e/\epsilon_0 \bigl( \rho_{\nu_b}^{B B^\prime} (\vec{r}) / \epsilon_\text{m} - \rho_{\nu_b}^{B B^\prime} (x,y,-z) / \epsilon_\text{eff}^\text{m-m} \bigr)$
\begin{equation}
	\phi_{\nu_b}^{B B^\prime} (\vec{r}) = - \int \text{d}^3r^\prime \; e G^\text{m-m} (\vec{r}, \vec{r}^\prime) \rho_{\nu_b}^{B B^\prime} (\vec{r}^\prime)
\end{equation}
of molecule $\nu_b$ and approximate it by the potential generated by point charges, the atomic partial charges $q_j^{B B^\prime}$ at the atomic positions $\vec{R}_{j_{\nu_b}}$:
\begin{equation} \label{eq:ESP_mol}
	\phi_{\nu_b}^{B B^\prime} (\vec{r}) \approx \sum_{j} G^\text{m-m} (\vec{r}, \vec{R}_{j_{\nu_b}}) q_j^{B B^\prime},
\end{equation}
where $\vec{R}_{j_{\nu_b}} = \vec{R}_{\nu_b} + \vec{r}_{j}$ is the sum of position $\vec{R}_{\nu_b}$ of molecule $\nu_b$ and the position $\vec{r}_{j}$ of the $j$-th atom of molecule $\nu_b$ relative to $\vec{R}_{\nu_b}$. Assuming identical, uniformly oriented molecules, these relative positions $\vec{r}_{j}$ and charges $q_i^{AB, \nu} \equiv q_i^{AB}$ are equal for all molecules. After introducing the same procedure of generating partial charges for molecule $\nu_a$, the Coulomb coupling is expressed as the electrostatic interaction between atomic partial charges \cite{Madjet:JPhysChemB:06}:
\begin{equation} \label{eq:m-mPartialChargeCouplingElement}
	\V{A, \nu_a}{B, \nu_b}{A^\prime, \nu_a}{B^\prime, \nu_b} \approx \sum_{i, j} G^\text{m-m} (\vec{R}_{i_{\nu_a}}, \vec{R}_{j_{\nu_b}}) q_i^{A A^\prime} q_j^{B B^\prime}.
\end{equation}

The concept of the partial charge approximation for intermolecular Coulomb matrix elements can be extended to the interfacial molecule-semiconductor coupling. Therefore, we define the one-particle density of the semiconductor substrate
$ \rho_{\vec{k} \vec{k}^\prime}^{\lambda \lambda^\prime} (\vec{r}) = \psi_{\lambda, \vec{k}}^* (\vec{r}) \psi_{\lambda^\prime, \vec{k}^\prime}^{\phantom{*}} (\vec{r}) $.
We introduce an electrostatic potential for the $n$-th unit cell of the semiconductor substrate and approximate it by the potential of the partial charges $q_i^{\lambda \lambda^\prime}$ at the relative positions $\vec{r}_i$ within one unit cell:
\begin{align} \label{eq:ESP_s}
	\begin{split}
		\phi_n^{\lambda \lambda^\prime, \vec{k} \vec{k}^\prime} & (\vec{r}) = - \frac{1}{A_\text{uc}} \int_{\text{UC}_n} \text{d}^3 r^\prime \; e G^\text{m-s} (\vec{r}, \vec{R}_i + \vec{r}^\prime) \, u_{\lambda, \vec{k}}^* (\vec{r}^\prime) \\
		& \times u_{\lambda^\prime, \vec{k}^\prime}^{\phantom{*}} (\vec{r}^\prime) \e^{i (\vec{k}^\prime - \vec{k}) \cdot \vec{r}^\prime_\parallel} \; \xi_\lambda^* (Z_n + z^\prime) \xi_{\lambda^\prime} (Z_n + z^\prime)
	\end{split} \\
	& \approx \sum_{i} G^\text{m-s} (\vec{r}, \vec{r}_{i}) q_i^{\lambda \lambda^\prime},
\end{align}
where $A_\text{uc}$ denotes the area of the ZnO unit cell, $\vec{R}_n$ the lattice vector connected to the $n$-th unit cell with $z$ component $Z_n$, $u_{\lambda, \vec{k}} (\vec{r})$ the lattice-periodic Bloch function, and $\xi_\lambda^* (z)$ the QW envelope function in $z$ direction. Note that we neglect the momentum dependence of the semiconductor partial charges and take the value at the $\Gamma$ point: $q_i^{\lambda \lambda^\prime, \vec{k} \vec{k}^\prime} \approx q_i^{\lambda \lambda^\prime, \vec{0} \vec{0}} \equiv q_i^{\lambda \lambda^\prime}$. This approximation is valid, since we consider only electronic states close to the band edges (see Sec.~\ref{sec:TransferRates}).

To rewrite the matrix element in terms of the approximative partial charges, the integral over $\vec{r}$ in Eq.~\eqref{eq:m-sCouplingElement} is transformed into a sum of integrals over the single unit cells and the invariance of the Bloch functions under a lattice translation is used:
\begin{align}
	\begin{split} \label{eq:m-sPartialChargeCouplingElement}
		\V{\lambda, \vec{k}}{A, \nu}{\lambda^\prime, \vec{k}^\prime}{B, \nu} & \approx \frac{1}{N_\text{uc}} \sum_{n=1}^{N_{\text{uc}}} \e^{i(\vec{k}^\prime - \vec{k}) \cdot {\vec{R}_n}_\parallel} \\
		& \times \sum_{i, j} G^\text{m-s} (\vec{R}_{\nu} + \vec{r}_j, \vec{R}_n + \vec{r}_i) q_i^{\lambda \lambda^\prime} q_j^{A B}.
	\end{split}
\end{align}
$N_\text{uc}$ is the total number of unit cells in the QW.

In this way, the complex field distribution of the molecules and the semiconductor outside the van der Waals radius of the atoms is represented by point charges at the atomic positions.
The partial charge technique can be applied, if the electrostatic potentials are known, e.g., from DFT calculations. A detailed description of how the partial charges are calculated in FHI-aims is given in Appendix~\ref{App1:PartialCharges}.
The partial charges obtained by fitting to the electrostatic potential from a DFT calculation give direct excess to the effective transition dipole moment, by summing over the charges at the atomic positions
\begin{equation} \label{eq:totalDipoleMomentS}
	\vec{d}_{\cb \vb} =\sum_i q_i^{\cb \vb}\vec{r}_i.
\end{equation}
This works analogously for the effective dipole moment of the L4P molecule:
\begin{equation} \label{eq:totalDipoleMomentM}
	\vec{d}_{\Lu \Ho} =\sum_j q_j^{\Lu \Ho}\vec{r}_j.
\end{equation}

The \textit{full} electrostatic potential is required for calculating the partial charges $q_i^{\lambda \lambda}$ and $q_j^{AA}$ that enter the monopole-monopole coupling elements. It is typically calculated by solving the Poisson equation for the full density. In FHI-aims a very efficient algorithm is used, that employs the multipole moments of the density \cite{Delley:JChemPhys:90,Delley:JComputChem:96}.
To approximate the F\"orster-type coupling elements, we need to calculate the \textit{transition} partial charges $q_i^{\lambda \lambda^\prime}$ ($\lambda \neq \lambda^\prime$) and $q_j^{AB}$ ($A \neq B$) for the involved electronic states (orbitals) or rather products of states. They can be calculated analogous to the charges calculated for the full potential, once the transition potential was calculated. Instead of using the full charge density as input for this algorithm, we implemented the option to calculate a transition density. The charges are fitted to the electrostatic potential calculated on different grids, based on these transition densities. The potential obtained from a DFT calculation employing the hybrid xc-functional HSE06 \cite{Heyd:JChemPhys:03} and the potential reconstructed from the partial charges is shown in Fig.~\ref{fig:PotentialFit}.

\subsection{Transformation of molecular orbitals into a Bloch basis} \label{sec:Transformation}

\begin{figure}
	\centering
	\includegraphics[width=0.35\linewidth]{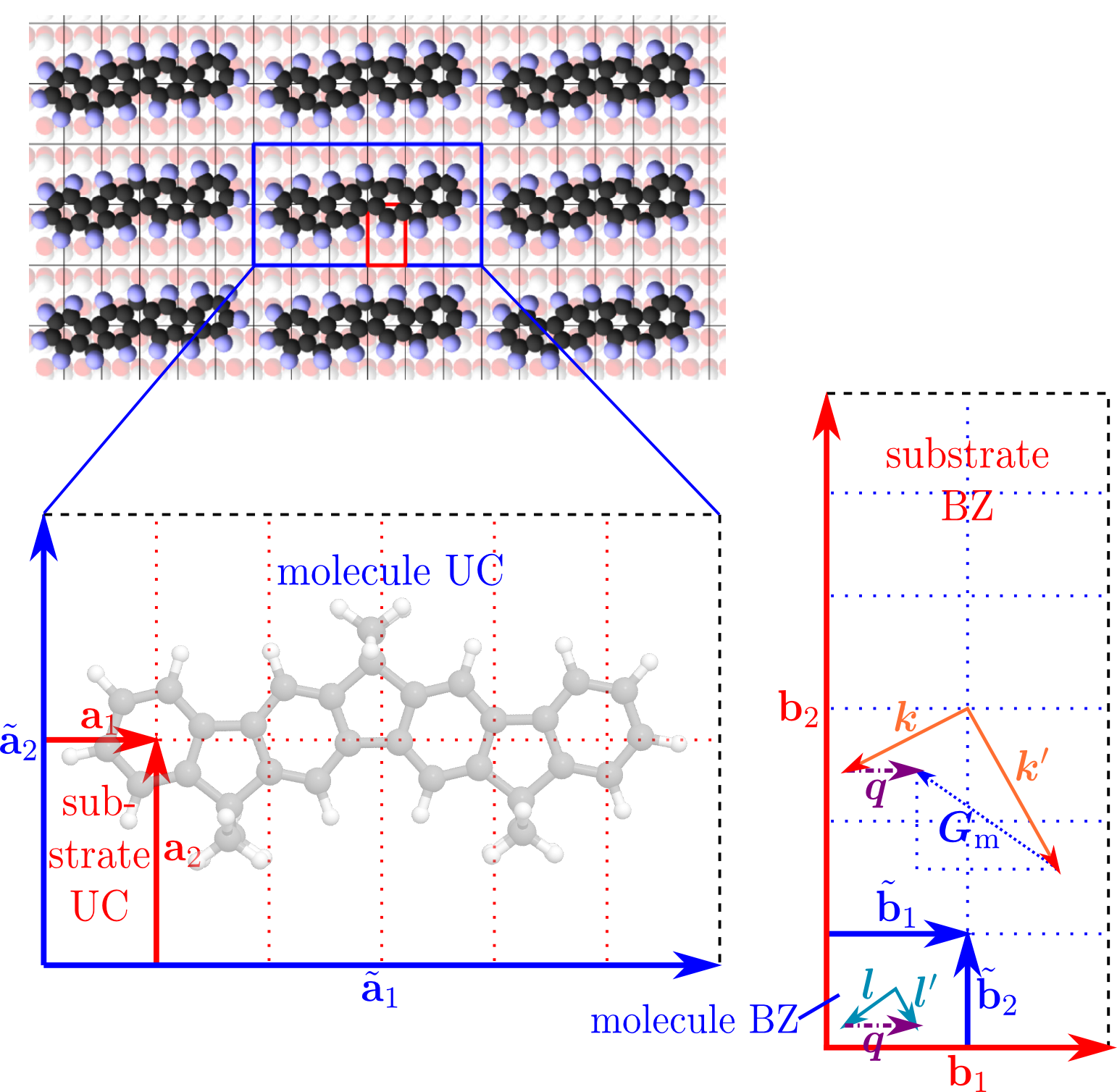}
	\caption{Unit cells (UCs, left) and Brillouin zones (BZs, right) of the hybrid system with the maximum coverage of 1 molecule per 12 substrate unit cells ($6 \times 2$) (cf. Refs.~\onlinecite{Verdenhalven:PhysRevB:14,Specht:ProcSPIE:16}). The momentum vector diagrams depicted in the substrate and molecule BZs illustrate the microscopic momentum selection rules of the interlayer coupling.}
	\label{fig:UCs_BZs_6x2}
\end{figure}

Following Ref.~\onlinecite{Verdenhalven:PhysRevB:14}, we assume a lattice-periodic arrangement of the organic molecules. Moreover, we assume that the substrate unit cells match the molecule unit cell such that the molecular lattice vectors are integer multiples of the substrate lattice vector, as illustrated in Fig.~\ref{fig:UCs_BZs_6x2}. This assumption leads to an idealized model used for the microscopic description. An extension to a disordered molecular layer is planned for future work \cite{Malic:PhysRevLett:11}.
We transform the molecular operators into a Bloch basis using
$ \hat{a}_{A, \nu} = 1 / \sqrt{N_\text{m}} \sum_{\vec{l}} \e^{-i \vec{l} \cdot {\vec{R}_\nu}_\parallel} \hat{a}_{A, \vec{l}}$ (cf. Refs.~\onlinecite{Slater:RevModPhys:34,Verdenhalven:PhysRevB:14}),
where we introduced the two-dimensional wave vectors $\vec{l}$ for the molecular states. The wave vectors $\vec{l}$ are restricted to the first Brillouin zone of the molecules, cf. Fig.~\ref{fig:UCs_BZs_6x2}. $N_\text{m}$ is the total number of molecules.
For a sufficiently extended molecular layer, we approximate \cite{Verdenhalven:PhysRevB:14}
\begin{equation}
	\sum_{\nu} \frac{1}{N_\text{m}} \e^{i \vec{Q} \cdot {\vec{R}_\nu}_\parallel} \approx
	\sum_{m_1, m_2 \in \mathbb{Z}} \delta_{\vec{Q}, m_1 \tilde{\vec{b}}_1 + m_2 \tilde{\vec{b}}_2} \equiv \sum_{\vec{G}_\text{m}} \delta_{\vec{Q}, \vec{G}_\text{m}}
\end{equation}
with $\vec{G}_\text{m} = m_1 \tilde{\vec{b}}_1 + m_2 \tilde{\vec{b}}_2$  a lattice vector in molecular reciprocal space (cf. Fig.~\ref{fig:UCs_BZs_6x2}).
The molecular free electron part in the new basis reads
\begin{align}
	\hat{H}_0^\text{m} = \sum_{A} \varepsilon_{A} \, \sum_{\vec{l}} \hat{a}_{A, \vec{l}}^\dagger \hat{a}_{A, \vec{l}}^{\phantom{\dagger}}.
\end{align}
The intermolecular Coulomb Hamiltonian in the momentum basis has the form
\begin{align}
	\begin{split} \label{eq:m-mCoulombHamiltonian}
		\hat{H}_{\text{C}}^{\text{m-m}} = & \frac{1}{2} \frac{1}{N_\text{m}} \sum_{A, B} \sum_{\vec{l}_1, \dots \vec{l}_4} \sum_{\vec{G}_\text{m}}  \delta_{\vec{l}_1 - \vec{l}_4, \vec{l}_3 - \vec{l}_2 + \vec{G}_\text{m}} \mathcalV{A}{B}{A}{B} (\vec{l}_2 - \vec{l}_3) \\
		& \times \hat{a}_{A, \vec{l}_1}^{\dagger} \hat{a}_{B, \vec{l}_2}^{\dagger} \hat{a}_{B, \vec{l}_3}^{\phantom{\dagger}} \hat{a}_{A, \vec{l}_4}^{\phantom{\dagger}}\\
		+ &
		\frac{1}{N_\text{m}} \sum_{\vec{l}_1, \dots, \vec{l}_4} \sum_{\vec{G}_\text{m}} \delta_{\vec{l}_1 - \vec{l}_4, \vec{l}_3 - \vec{l}_2 + \vec{G}_\text{m}} \mathcalV{\Ho}{\Lu}{\Lu}{\Ho} (\vec{l}_2 - \vec{l}_3) \\
		& \times \hat{a}_{\Ho, \vec{l}_1}^{\dagger} \hat{a}_{\Lu, \vec{l}_2}^{\dagger} \hat{a}_{\Ho, \vec{l}_3}^{\phantom{\dagger}} \hat{a}_{\Lu, \vec{l}_4}^{\phantom{\dagger}},
	\end{split}
\end{align}
with the coupling element using partial charges
\begin{align}
	\begin{split}
		\mathcalV{A}{B}{A^\prime}{B^\prime} (\vec{q}) = & \sum_{\vec{\Delta}_{\text{m-m}} \neq \vec{0}} \e^{i \vec{q} \cdot {\vec{\Delta}_{\text{m-m}}}_\parallel}
		\sum_{i, j} q_i^{A A^\prime} q_j^{B B^\prime} \\
		\times & G^\text{m-m} (\vec{r}_i, \vec{r}_{j} + \vec{\Delta}_{\text{m-m}}).
	\end{split}
\end{align}
Here, the sum over $\vec{\Delta}_{\text{m-m}} \equiv \vec{R}_{\nu_b} - \vec{R}_{\nu_a}$ runs over all difference vectors between the positions of two molecular unit cells (with one cell at a fixed position). The Kronecker delta in Eq.~\eqref{eq:m-mCoulombHamiltonian} ensures momentum conservation except for a reciprocal lattice vector.
The interlayer Coulomb Hamiltonian is also transformed into momentum representation:
\begin{align}
	\begin{split} \label{eq:m-sHamiltonianMomentumSpace}
		\hat{H}_{\text{C}}^{\text{m-s}} = &  \frac{1}{N_\text{uc}} \sum_{\lambda, \vec{k},  \vec{k}^\prime} \sum_{A, \vec{l}, \vec{l}^\prime} \sum_{\vec{G}_\text{m}} \delta_{\vec{l} - \vec{l}^\prime, \vec{k} - \vec{k}^\prime + \vec{G}_\text{m}} \mathcalV{\lambda}{A}{\lambda}{A} (\vec{k}^\prime - \vec{k}) \\
		& \times \hat{a}_{\lambda, \vec{k}}^{\dagger} \hat{a}_{A, \vec{l}}^{\dagger} \hat{a}_{A, \vec{l}^\prime}^{\phantom{\dagger}} \hat{a}_{\lambda, \vec{k}^\prime}^{\phantom{\dagger}} \\
		& + \frac{1}{N_\text{uc}} \Biggl( \sum_{\vec{k}, \vec{k}^\prime} \sum_{\vec{l}, \vec{l}^\prime} \sum_{\vec{G}_\text{m}} \delta_{\vec{l} - \vec{l}^\prime, \vec{k} - \vec{k}^\prime + \vec{G}_\text{m}}
		\mathcalV{\cb}{\Ho}{\vb}{\Lu} (\vec{k}^\prime - \vec{k}) \\
		& \times \hat{a}_{\cb, \vec{k}}^{\dagger} \hat{a}_{\Ho, \vec{l}}^{\dagger} \hat{a}_{\Lu, \vec{l}^\prime}^{\phantom{\dagger}} \hat{a}_{\vb, \vec{k}^\prime}^{\phantom{\dagger}} + \textit{h.c.} \Biggr)
	\end{split}
\end{align}
with the redefined matrix elements
\begin{align}
	\begin{split}
		\mathcalV{\lambda}{A}{\lambda^\prime}{A^\prime} (\vec{q}) = \sum_{\vec{\Delta}_\text{m-s}} \e^{i \vec{q} \cdot {\vec{\Delta}_\text{m-s}}_\parallel}
		\sum_{i, j} q_i^{\lambda \lambda^\prime} q_j^{A A^\prime} G^\text{m-s} (\vec{r}_j, \vec{r}_i + \vec{\Delta}_\text{m-s}).
	\end{split}
\end{align}
The sum over $\vec{\Delta}_\text{m-s} \equiv \vec{R}_i - \vec{R}_\nu$ runs over the positions $\vec{R}_i$ of all substrate unit cells relative to a fixed molecular position $\vec{R}_\nu$. The Kronecker $\delta$ in Eq.~\eqref{eq:m-sHamiltonianMomentumSpace} ensures quasi-momentum conservation during F\"orster transfer and imposes microscopic momentum selection rules on the system. The change in momentum in the molecule has to match the momentum change in the semiconductor substrate except for a reciprocal lattice vector $\vec{G}_\text{m}$ of the molecules, cf. Fig.~\ref{fig:UCs_BZs_6x2} (right panel).

\begin{figure}
	\centering
	\includegraphics[width=0.475\linewidth]{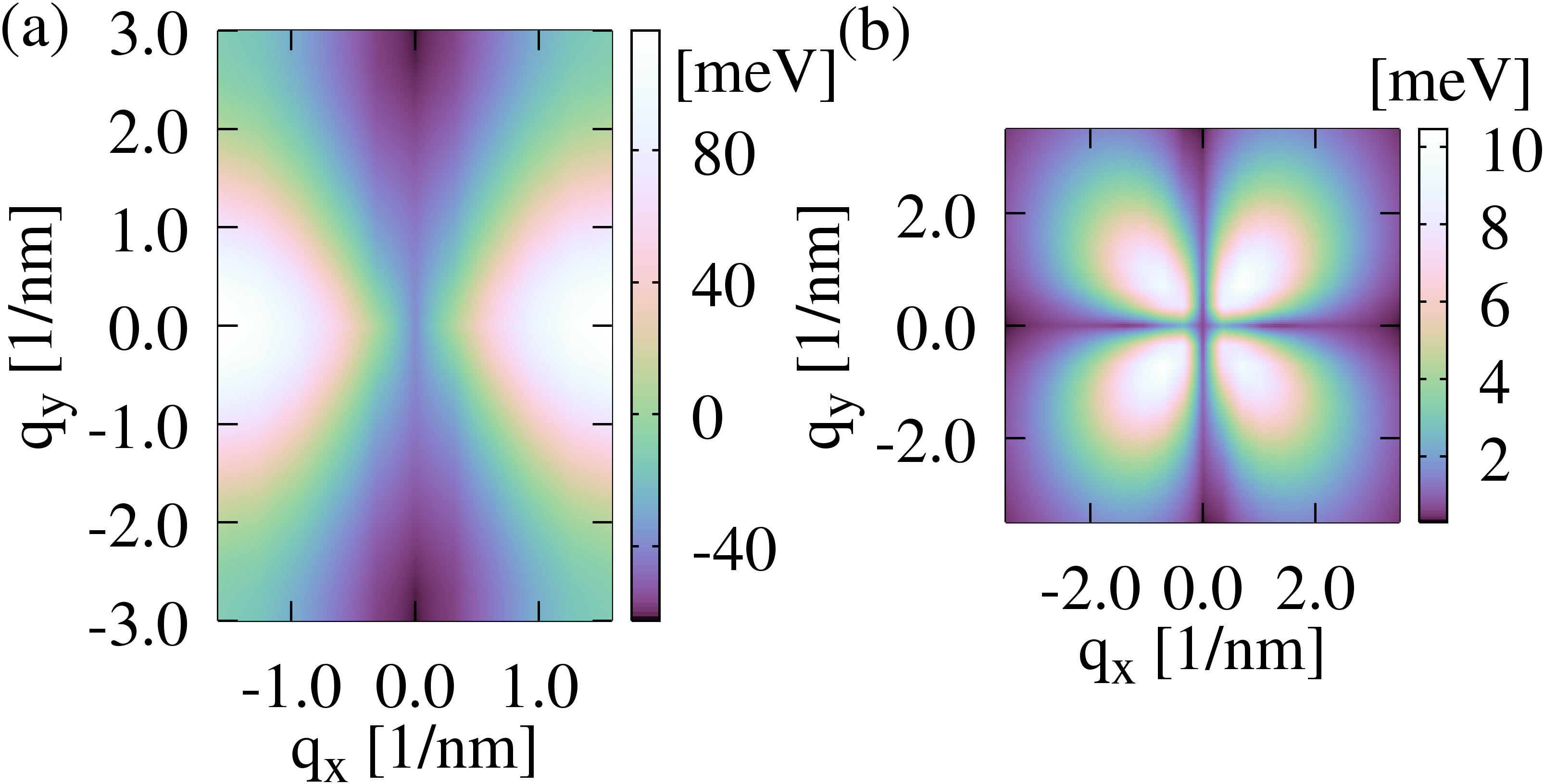}
	\caption{(a) Contour plot showing the magnitude of the intermolecular F\"orster coupling element $\mathcalV{\Ho}{\Lu}{\Lu}{\Ho} (\vec{q})$ in $\unit{meV}$ over all $\vec{q}$ values belonging to the first molecule Brillouin zone for the maximum coverage of 1 molecule per 12 substrate unit cells ($6 \times 2$). (b) Contour plot for the coupling strength $\lvert \mathcalV{\cb}{\Ho}{\vb}{\Lu} (\vec{q}) \rvert$ of the interlayer F\"orster interaction in $\unit{meV}$.}
	\label{fig:FoersterCouplingElement}
\end{figure}

In Fig.~\ref{fig:FoersterCouplingElement} (a), the intermolecular F\"orster coupling element $\mathcalV{\Ho}{\Lu}{\Lu}{\Ho} (\vec{q})$ is plotted. It exhibits a dumbbell-like shape along the $x$ direction, since the effective dipole moments (given in Eq.~\eqref{eq:totalDipoleMomentM}) of all uniformly oriented molecules in the organic film point into the $x$ direction, such that the maxima of the coupling strength also lie along the $x$ axis \cite{Verdenhalven:PhysRevB:14}.
Fig.~\ref{fig:FoersterCouplingElement} (b) shows the interlayer F\"orster coupling strength $\lvert \mathcalV{\cb}{\Ho}{\vb}{\Lu} (\vec{q}) \rvert$. The shape of the transfer element shows four lobes oriented roughly along the diagonals, since the effective dipole moments obtained from the transition partial charges (cf. Eqs.~\eqref{eq:totalDipoleMomentS} and \eqref{eq:totalDipoleMomentM}) are oriented almost perpendicular to each other along the coordinate axes: $\vec{d}_{\Lu \Ho} \parallel \vec{e}_x$, $\vec{d}_{\cb \vb} \parallel \vec{e}_y$.

\subsection{Molecular exciton basis} \label{sec:MolecularExcitons}

The state $\ket{\phi_0^\text{m}}$ denotes the ground state of the organic layer where the HOMOs of all molecules are fully occupied. A basis describing exciton states is defined using the ground state and the annihilation (creation) operators $\hat{a}_{A, \vec{l}}^{(\dagger)}$ for electrons in the molecule:
\begin{equation}
	\ket{ \vec{l}_1, \vec{l}_2 } \equiv \hat{a}^\dagger_{\Lu, \vec{l}_1} \hat{a}^{\phantom{\dagger}}_{\Ho, \vec{l}_2} \ket{\phi_0^\text{m}}
\end{equation}
Due to the intermolecular Coulomb interaction, the coupled excitonic states of the molecular layer are delocalized superpositions of the two-particle states:
\begin{equation} \label{eq:molExciton}
	\ket{X_\alpha^\text{m}} = \sum_{\vec{l}_1, \vec{l}_2} c_{\vec{l}_1, \vec{l}_2}^\alpha \ket{ \vec{l}_1, \vec{l}_2 } 
	= \sum_{\vec{l}, \vec{q}} c_{\vec{l} + \vec{q}, \vec{l}}^\alpha \ket{\vec{l} + \vec{q}, \vec{l}}.
\end{equation}
Here, the wave vectors of the exciton basis are given using $\vec{l} \equiv \vec{l}_2$ and momentum transfer $\vec{q} \equiv \vec{l}_1 - \vec{l}_2$.

The eigenvalue problem for the molecular Frenkel exciton states $\ket{X_\alpha^\text{m}}$ is given by
$(\hat{H}^\text{m}_0 + \hat{H}^\text{m-m}_\text{C}) \ket{X_\alpha^\text{m}} = (E_0^\text{m} + E_\alpha^\text{m}) \ket{X_\alpha^\text{m}}$,
where we introduced the molecular eigenenergy $E_0^\text{m} + E_\alpha^\text{m}$ that solves the Schr\"odinger equation of the molecular layer.
$E_0^\text{m} = \bra{\phi_0^\text{m}} \hat{H}^\text{m}_0 + \hat{H}^\text{m-m}_\text{C} \ket{\phi_0^\text{m}}$ denotes the constant ground state energy.

A representation of the eigenproblem in the two-particle basis has the form
\begin{equation} \label{eq:molEigenproblem}
	\bra{\vec{l} + \vec{q}, \vec{l}} \hat{H}^\text{m}_0 + \hat{H}^\text{m-m}_\text{C} \ket{X_\alpha^\text{m}}
	= (E_0^\text{m} + E_\alpha^\text{m}) \; c_{\vec{l} + \vec{q}, \vec{l}}^\alpha,
\end{equation}
where the left-hand side of the equation depends on the coefficients $c_{\vec{l} + \vec{q}, \vec{l}}^\alpha$ according to Eq.~\eqref{eq:molExciton}.
Using the Hamiltonian in momentum representation (as derived in Sec.~\ref{sec:Transformation}), the left-hand side of this equation is evaluated numerically using discrete wave vectors $\vec{l}_i$, $\vec{q}_j$ (each having $N_\text{d}^\text{m}$ values). The calculation is shown in Appendix~\ref{App2:MolecularEigenproblem}. The full molecular eigenproblem is blockdiagonal with respect to $\vec{q}_j$, leading to an analytical solution for the eigenenergies $E_{\alpha = \vec{q}_j, n}^\text{m}$ and eigenvector components $c_{\vec{l}_i + \vec{q}_j, \vec{l}_i}^{\alpha = \vec{q}_j, n}$. Two solutions emerge:
The $N_\text{d}^\text{m} - 1$ excitonic basis states corresponding to the degenerate eigenvalue $E_-^\text{m}$ are pairwise antisymmetric linear combinations of the two-particle basis functions with equal momentum transfer $\vec{q}_j$:
\begin{equation} \label{eq:antisymmetricEigenvectors}
	\ket{X_{\vec{q}_j, n}^\text{m}} = \sqrt{\frac{N_\text{m}}{N_\text{d}^\text{m}}} \frac{1}{\sqrt{2}} \bigl( \ket{\vec{l}_1 + \vec{q}_j, \vec{l}_1} - \ket{\vec{l}_{n+1} + \vec{q}_j, \vec{l}_{n+1}} \bigr)
\end{equation}
for $n \in \{1, \dots, N_\text{d}^\text{m} - 1 \}$.
The non-degenerate eigenvalue $E_{\vec{q}_j +}^\text{m}$ belongs to a fully symmetric eigenvector:
\begin{equation} \label{eq:symmetricEigenvectors}
	\ket{X_{\vec{q}_j +}^\text{m}} = \frac{\sqrt{N_\text{m}}}{N_\text{d}^\text{m}} \sum_{i=1}^{N_\text{d}^\text{m}} \ket{\vec{l}_i + \vec{q}_j, \vec{l}_i}.
\end{equation}
It appears that, in contrast to $E_{\vec{q}_j +}^\text{m}$, the degenerate eigenenergy $E_-^\text{m}$ is dispersionless, i.e., does not depend on $\vec{q}_j$. Later it is shown that the corresponding antisymmetric eigenvectors form dark states, such that only the symmetric eigenstates contribute to the charge transfer across the hybrid interface.

\section{Equations of motion of the hybrid system} \label{sec:EOM}

\begin{figure}
	\centering
	\includegraphics[width=0.48\linewidth]{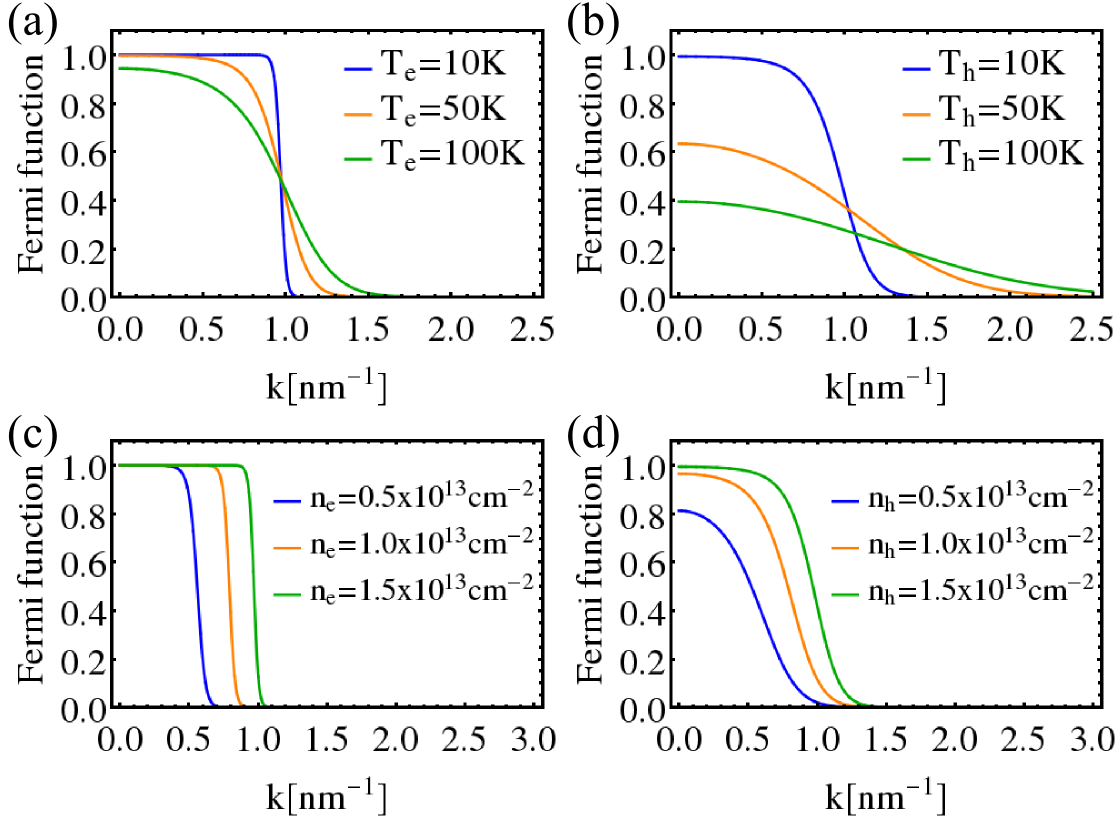}
	\caption{Fermi functions of electrons ((a) and (c)) and holes ((b) and (d)) for the parameter set given in Tab.~\ref{tab:parameters} and varying temperatures ((a) and (b)) and charge carrier concentrations ((c) and (d)).}
	\label{fig:Fermiplot}
\end{figure}

Now we focus on the transfer of semiconductor excitations to the molecules:
The equations of motion (EOMs) in the excitonic basis are derived using the von Neumann equation:
\begin{equation}
	i \hbar \frac{\partial}{\partial t} \trace_\text{s} \bigl[ \bra{a} \hat{O}_\text{s} \hat{\rho} \ket{b} \bigr] = \trace_\text{s} \bigl[ \bra{a} \hat{O}_\text{s} [\hat{H}, \hat{\rho}]_{-} \ket{b} \bigr]
\end{equation}
for the density operator  $\hat{\rho} \equiv \hat{\rho}_\text{m} \otimes \hat{\rho}_\text{s}$. $a, b$ are states of the molecular system and $\hat{O}_\text{s}$ is an operator of the semiconductor system or the identity.
In the following, we are interested in the population of the molecular system $\rho_{\vec{q}_j, n}^\text{m} \equiv \trace_\text{s} \bigl[ \bra{X_{\vec{q}_j, n}^{\text{m}}} \hat{\rho} \ket{X_{\vec{q}_j, n}^{\text{m}}} \bigr]$, assuming approximative spatial homogeneity in the molecular layer (identical $\vec{q}_j$ index). We introduce the assisted molecule--semiconductor coherence
$\sigma{\substack{\vec{k}, \vec{k}^\prime \\ \vec{q}_j, n}} \equiv \trace_\text{s} \bigl[ \hat{a}_{\cb, \vec{k}}^{\dagger} \hat{a}_{\vb, \vec{k}^\prime}^{\phantom{\dagger}} \bra{X_{\vec{q}_j, n}^{\text{m}}} \hat{\rho} \ket{\phi_0^\text{m}} \bigr]$.
We do not express the full system Hamiltonian $\hat{H}$ in the new basis, since it is sufficient to evaluate how the Hamiltonian acts on the new basis states.
The equation of motion for the exciton density is given by:
\begin{align}
	\begin{split}
		\frac{\partial}{\partial t} \rho_{\vec{q}_j, n}^\text{m} = &
		- \frac{2}{\hbar} \frac{1}{N_\text{uc}} \tilde{c}_{\vec{q}_j, n} \sum_{\vec{k}, \vec{k}^\prime} \sum_{\vec{G}_\text{m}} \delta_{\vec{q}_j, \vec{k}^\prime - \vec{k} + \vec{G}_\text{m}} \\
		& \times \im \left[ \mathcalV{\cb}{\Ho}{\vb}{\Lu} (\vec{k}^\prime - \vec{k}) \sigma{\substack{ \vec{k}, \vec{k}^\prime \\ \vec{q}_j, n}} \right],
	\end{split}
\end{align}
where we defined $\tilde{c}_{\vec{q}_j, n} \equiv \tfrac{N_\text{m}}{N_\text{d}^\text{m}} \sum_{i} c_{\vec{l}_i + \vec{q}_j, \vec{l}_i}^{\vec{q}_j, n}$. The Kronecker delta ensures momentum conservation during interlayer F\"orster transfer, cf. Fig.~\ref{fig:UCs_BZs_6x2}. Obviously, the molecule--semiconductor coherence between the layers is the source term of the molecular occupation.
$\tilde{c}_{\vec{q}_j, n} = \tfrac{1}{\sqrt{2}} - \tfrac{1}{\sqrt{2}}$ vanishes for the antisymmetric solution of the molecular eigenproblem (cf. Eq.~\eqref{eq:antisymmetricCoefficients} in Appendix~\ref{App2:MolecularEigenproblem}), such that only the symmetric (bright) states contribute:
\begin{align}
	\begin{split}
		\frac{\partial}{\partial t} \rho_{\vec{q}_j +}^\text{m} = &
		- \frac{2}{\hbar} \frac{\sqrt{N_\text{d}^\text{m}}}{N_\text{uc}} \sum_{\vec{k}, \vec{k}^\prime} \sum_{\vec{G}_\text{m}} \delta_{\vec{q}_j, \vec{k}^\prime - \vec{k} + \vec{G}_\text{m}} \\
		& \times \im \left[ \mathcalV{\cb}{\Ho}{\vb}{\Lu} (\vec{k}^\prime - \vec{k}) \sigma{\substack{\vec{k}, \vec{k}^\prime \\ \vec{q}_j +}} \right], \label{eq:DiagonalExcitonDensity} 
	\end{split} \\
	\begin{split} \label{eq:HomogeneousExcitonSubstrateCoherence}
		\frac{\partial}{\partial t} \sigma{\substack{\vec{k}, \vec{k}^\prime \\ \vec{q}_j +}} = &
		\frac{i}{\hbar} \bigl( \varepsilon^{\vec{k}}_{\cb} - \varepsilon^{\vec{k}^\prime}_{\vb} - E_{\vec{q}_j +}^\text{m} + \mathcal{V}_\text{mono}^\text{m-s} \bigr) \sigma{\substack{\vec{k}, \vec{k}^\prime \\ \vec{q}_j +}} \\
		& + \frac{i}{\hbar} \frac{\sqrt{N_\text{d}^\text{m}}}{N_\text{uc}} \sum_{\vec{G}_\text{m}} \delta_{\vec{q}_j, \vec{k}^\prime - \vec{k} + \vec{G}_\text{m}} {\mathcalV{\cb}{\Ho}{\vb}{\Lu}}^* (\vec{k}^\prime - \vec{k}) \\
		& \times \bigl( (1 - f_{\text{h}, \vec{k}^\prime}) (1 - f_{\text{e}, \vec{k}}) \rho_{\vec{q}_j +}^\text{m} - f_{\text{h}, \vec{k}^\prime} f_{\text{e}, \vec{k}} \rho_0^\text{m} \bigr)
	\end{split}
\end{align}
with $\rho_0^\text{m} \equiv \trace_\text{s} \bigl[ \bra{\phi_0^\text{m}} \hat{\rho} \ket{\phi_0^\text{m}} \bigr]$.
The constant shift
\begin{align}
	\begin{split}
		\mathcal{V}_\text{mono}^\text{m-s} \equiv & \frac{N_\text{m}}{N_\text{uc}} \bigl( \mathcalV{\cb}{\Ho}{\cb}{\Ho} (\vec{0}) - \mathcalV{\vb}{\Ho}{\vb}{\Ho} (\vec{0}) \bigr) \\
		& + ( 1 -  \tfrac{1}{2} n_\text{h}^\text{2D} A_\text{uc} ) \bigl( \mathcalV{\vb}{\Ho}{\vb}{\Ho} (\vec{0}) - \mathcalV{\vb}{\Lu}{\vb}{\Lu} (\vec{0}) \bigr) \\
		& + \tfrac{1}{2} n_\text{e}^\text{2D} A_\text{uc} \bigl( \mathcalV{\cb}{\Ho}{\cb}{\Ho} (\vec{0}) - \mathcalV{\cb}{\Lu}{\cb}{\Lu} (\vec{0}) \bigr)
	\end{split}
\end{align}
represents the monopole-monopole interaction, where $n_{\text{e/h}}^\text{2D} = N_{\text{e/h}} / A_\text{QW}$ is the two-dimensional carrier density for electrons (e) and holes (h). It describes the self-energy due to the electrostatic coupling of the electronic states in the molecular layer and the semiconductor substrate. For the derivation of the above equations of motion, spatial homogeneity for fixed semiconductor populations was assumed using $\langle \hat{a}_{\lambda, \vec{k}}^{\dagger} \hat{a}_{\lambda, \vec{k}^\prime}^{\phantom{\dagger}} \rangle = \delta_{\vec{k}, \vec{k}^\prime} \langle \hat{a}_{\lambda, \vec{k}}^{\dagger} \hat{a}_{\lambda, \vec{k}}^{\phantom{\dagger}} \rangle$. As a consequence, inhomogeneous monopole-monopole contributions were neglected. A Hartree-Fock factorization was applied to the semiconductor part and we set $\delta_{\vec{k}-\vec{k}^\prime, \vec{G}_\text{m}} = \delta_{\vec{k}, \vec{k}^\prime} \delta_{\vec{G}_\text{m}, \vec{0}}$, since only $\vec{k}$ states close to the $\Gamma$ point are relevant. Only a single excitation in the molecular layer is considered and the coherences $\langle \hat{a}_{\vb, \vec{k}}^{\dagger} \hat{a}_{\cb, \vec{k}^\prime}^{\phantom{\dagger}} \rangle$ are assumed to decay rapidly, such that $\langle \hat{a}_{\vb, \vec{k}_1}^{\dagger} \hat{a}_{\cb, \vec{k}_2}^{\phantom{\dagger}} \rangle \langle \hat{a}_{\cb, \vec{k}_3}^{\dagger} \hat{a}_{\vb, \vec{k}_4}^{\phantom{\dagger}} \rangle \approx 0$ Moreover, the system was assumed to be in the thermodynamic quasi-equilibrium within the valence and conduction band, such that the subband carrier populations are described by Fermi distribution functions,
$f_{i, \vec{k}} = \left( \exp \left( (\varepsilon_i^{\vec{k}} - \mu_i)/(k_\text{B} T_i) \right) + 1 \right)^{-1}$ with $i = \text{e, h}$ for electrons and holes \cite{Haug::04,Chow::12} (cf. Fig.~\ref{fig:Fermiplot}).
$\mu_{i}$ denotes the carrier quasi-equilibrium chemical potential in the respective band and $T_i$ is the non-equilibrium temperature, which can be different for electrons and holes.
We focus on situations where only states close to the $\Gamma$ point ($\vec{k} = \vec{0}$) are populated. Therefore, a description of the band structure around the $\Gamma$ point using the effective-mass approximation is possible with effective masses $m_\vb^* < 0$ and $m_\cb^* > 0$ and
$ \varepsilon_{\text{h}}^{\vec{k}} = - (\hbar^2 k^2)/(2 m_\vb^*) = - \varepsilon_\vb^{\vec{k}}$,
$\varepsilon^{\vec{k}}_{\text{e}} = (\hbar^2 k^2)/(2 m_\cb^*) = \varepsilon_\cb^{\vec{k}} - \varepsilon_\text{gap}^\text{s}$.
The effective masses $m_\vb^* =\unit[-8.3035]{m_0} $ of the valence and $m_\cb^* =\unit[1.4463]{m_0}$ of the conduction band electrons are obtained from a fit to the DFT band structure of ZnO surface bands \cite{Verdenhalven:PhysRevB:14}.
The chemical potential for each band of a two-band system is calculated using \cite{Haug::04,Chow::99}:
\begin{equation} \label{eq:chemicalPotential}
	\mu_i = k_\text{B} T_i \ln \left( \exp \left( \frac{\pi \hbar^2 n_i^\text{2D}}{ m_i k_\text{B} T_i} \right) - 1 \right).
\end{equation}

The electron-hole states in the semiconductor substrate form a continuum, thus allowing to solve Eq.~\eqref{eq:HomogeneousExcitonSubstrateCoherence} for the assisted molecular exciton--substrate polarization $\sigma{\substack{\vec{k}, \vec{k}^\prime \\ \vec{q}_j +}}$ in the Markov approximation:
\begin{align}
	\begin{split} \label{eq:MarkovPolarization}
		\sigma{\substack{\vec{k}, \vec{k}^\prime \\ \vec{q}_j +}} = &
		- i \pi \frac{\sqrt{N_\text{d}^\text{m}}}{N_\text{uc}} \sum_{\vec{G}_\text{m}} \delta_{\vec{q}_j, \vec{k}^\prime - \vec{k} + \vec{G}_\text{m}} {\mathcalV{\cb}{\Ho}{\vb}{\Lu}}^* (\vec{k}^\prime - \vec{k}) \\
		& \times \bigl( f_{\text{h}, \vec{k}^\prime} f_{\text{e}, \vec{k}} \rho_0^\text{m} - (1 - f_{\text{h}, \vec{k}^\prime}) (1 - f_{\text{e}, \vec{k}}) \rho_{\vec{q}_j +}^\text{m} \bigr) \\ 
		& \times \delta \left( \varepsilon_\text{e}^{\vec{k}} + \varepsilon_\text{h}^{\vec{k}^\prime} - \Delta_{\vec{q}_j} \right),
	\end{split}
\end{align}
where we introduced $\Delta_{\vec{q}_j} \equiv E_{\vec{q}_j +}^\text{m} - \varepsilon_\text{gap}^\text{s} - \mathcal{V}_\text{mono}^\text{m-s}$.
This solution is inserted into the equation of motion for the homogeneous molecule density $\rho_{\vec{q}_j +}^\text{m}$ (Eq.~\eqref{eq:DiagonalExcitonDensity}):
\begin{align}
	\begin{split} \label{eq:EOMrhom}
		\frac{\partial}{\partial t} \rho_{\vec{q}_j +}^\text{m} = & \frac{2 \pi}{\hbar} \frac{N_\text{d}^\text{m}}{N_\text{uc}^2} \sum_{\vec{k}, \vec{k}^\prime} \sum_{\vec{G}_\text{m}} \delta_{\vec{q}_j, \vec{k}^\prime - \vec{k} + \vec{G}_\text{m}} \left\lvert \mathcalV{\cb}{\Ho}{\vb}{\Lu} (\vec{k}^\prime - \vec{k}) \right\rvert^2 \\
		& \times \bigl( f_{\text{h}, \vec{k}^\prime} f_{\text{e}, \vec{k}} \rho_0^\text{m} - (1 - f_{\text{h}, \vec{k}^\prime}) (1 - f_{\text{e}, \vec{k}}) \rho_{\vec{q}_j +}^\text{m} \bigr) \\
		& \times \delta \left( \varepsilon_\text{e}^{\vec{k}} + \varepsilon_\text{h}^{\vec{k}^\prime} - \Delta_{\vec{q}_j} \right).
	\end{split}
\end{align}
This equation of motion for the transfer from the semiconductor electron-hole continuum to the molecular system allows to derive microscopical rate equations for (Coulomb) scattering processes in the heterostructure similar to Refs.~\onlinecite{Wolters:PhysRevB:09,Malic:NewJournalofPhysics:10}.

\section{Discussion of the interlayer transfer rate} \label{sec:TransferRates}

From the equation of motion for the population $\rho_{\vec{q}_j +}^\text{m} = \trace_\text{s} \bigl[ \bra{X_{\vec{q}_j +}^{\text{m}}} \hat{\rho} \ket{X_{\vec{q}_j +}^{\text{m}}} \bigr]$ given in Eq.~\eqref{eq:EOMrhom}, the in-scattering rate
\begin{align}
	\begin{split} \label{eq:TransferRate}
		\Gamma_{\vec{q}_j +}^\text{in} = & \frac{2 \pi}{\hbar} \frac{N_\text{d}^\text{m}}{N_\text{uc}^2} \sum_{\vec{k}, \vec{k}^\prime} \sum_{\vec{G}_\text{m}} \delta_{\vec{q}_j, \vec{k}^\prime - \vec{k} + \vec{G}_\text{m}} \left\lvert \mathcalV{\cb}{\Ho}{\vb}{\Lu} (\vec{k}^\prime - \vec{k}) \right\rvert^2 \\
		& \times f_{\text{h}, \vec{k}^\prime} f_{\text{e}, \vec{k}}
		\; \delta \left( \varepsilon_\text{e}^{\vec{k}} + \varepsilon_\text{h}^{\vec{k}^\prime} - \Delta_{\vec{q}_j} \right)
	\end{split}
\end{align}
is identified as transfer rate from the semiconductor substrate into the exciton state $X_{\vec{q}_j}^{\text{m}+}$ of the molecular layer. It is determined by the interlayer F\"orster coupling strength, the Fermi functions $f_{\text{h}, \vec{k}^\prime} f_{\text{e}, \vec{k}}$ representing the quasi-equilibrium carrier distributions in the QW, and the momentum and energy conservation.
In the same way, the back-scattering into the semiconductor layer is determined by:
\begin{align}
	\begin{split} \label{eq:TransferRate_bs}
		\Gamma_{\vec{q}_j +}^\text{out} = & \frac{2 \pi}{\hbar} \frac{N_\text{d}^\text{m}}{N_\text{uc}^2} \sum_{\vec{k}, \vec{k}^\prime} \sum_{\vec{G}_\text{m}} \delta_{\vec{q}_j, \vec{k}^\prime - \vec{k} + \vec{G}_\text{m}} \left\lvert \mathcalV{\cb}{\Ho}{\vb}{\Lu} (\vec{k}^\prime - \vec{k}) \right\rvert^2 \\
		& \times (1 - f_{\text{h}, \vec{k}^\prime}) (1 - f_{\text{e}, \vec{k}})
		\delta \left( \varepsilon_\text{e}^{\vec{k}} + \varepsilon_\text{h}^{\vec{k}^\prime} - \Delta_{\vec{q}_j} \right)
	\end{split}
\end{align}
with the typical Pauli blocking terms preventing back-scattering into the substrate when the relevant states are already occupied.
To obtain the total transfer rates involving all molecular exciton states, we sum over all numerically discrete $\vec{q}_j$ vectors within the first Brillouin zone of the molecules.
The total rate scales linearly with the total number of molecules $N_\text{m}$ in the system. To numerically evaluate the rate referring to one molecule (mean scattering between the inorganic semiconductor QW and one molecule of the organic layer), we calculate $\Gamma_\text{tot}^\text{in/out} / N_\text{m}$.

\begin{table}[b]
	\begin{tabular}{l l l}
		\hline\hline
		L4P relative permittivity & $\epsilon_\text{m}$ & $1.0$ \\
		ZnO relative permittivity \cite{Yoshikawa:JpnJApplPhys:97} & $\epsilon_\text{s}$ & $7.9$ \\
		ZnO band gap \cite{Yoshikawa:JpnJApplPhys:97} & $\varepsilon_\text{gap}^\text{s}$ & $\unit[3.4]{eV}$ \\
		2D electron density in ZnO & $n_\text{e}^\text{2D}$ & $\unit[1.5 \times 10^{13}]{/cm^2}$ \\
		2D hole density in ZnO & $n_\text{h}^\text{2D}$ & $\unit[1.5 \times 10^{13}]{/cm^2}$ \\
		Electron temperature in ZnO & $T_\text{e}$ & $\unit[10]{K}$ \\
		Hole temperature in ZnO & $T_\text{h}$ & $\unit[10]{K}$ \\
		Interlayer separation & $\Delta z$ & $\unit[0.4]{nm}$ \\ 
		Molecular coverage & & $10 \times 10$ unit cells$^2$\\
		Detuning & $\Delta_{\vec{0}}$ & \unit[15]{meV} \\
		\hline\hline
	\end{tabular}
	\caption{Material parameters used for calculating the HIOS transfer rates (if not varied in the plots).}
	\label{tab:parameters}
\end{table} 
For our analysis, while a parameter is varied the other material parameters are set as given in Tab.~\ref{tab:parameters}.

\subsection{Changing the detuning} \label{sec:Detuning}

\begin{figure}
	\hfill
	\begin{minipage}[b]{0.2\linewidth}
		\centering
		\includegraphics[width=\linewidth]{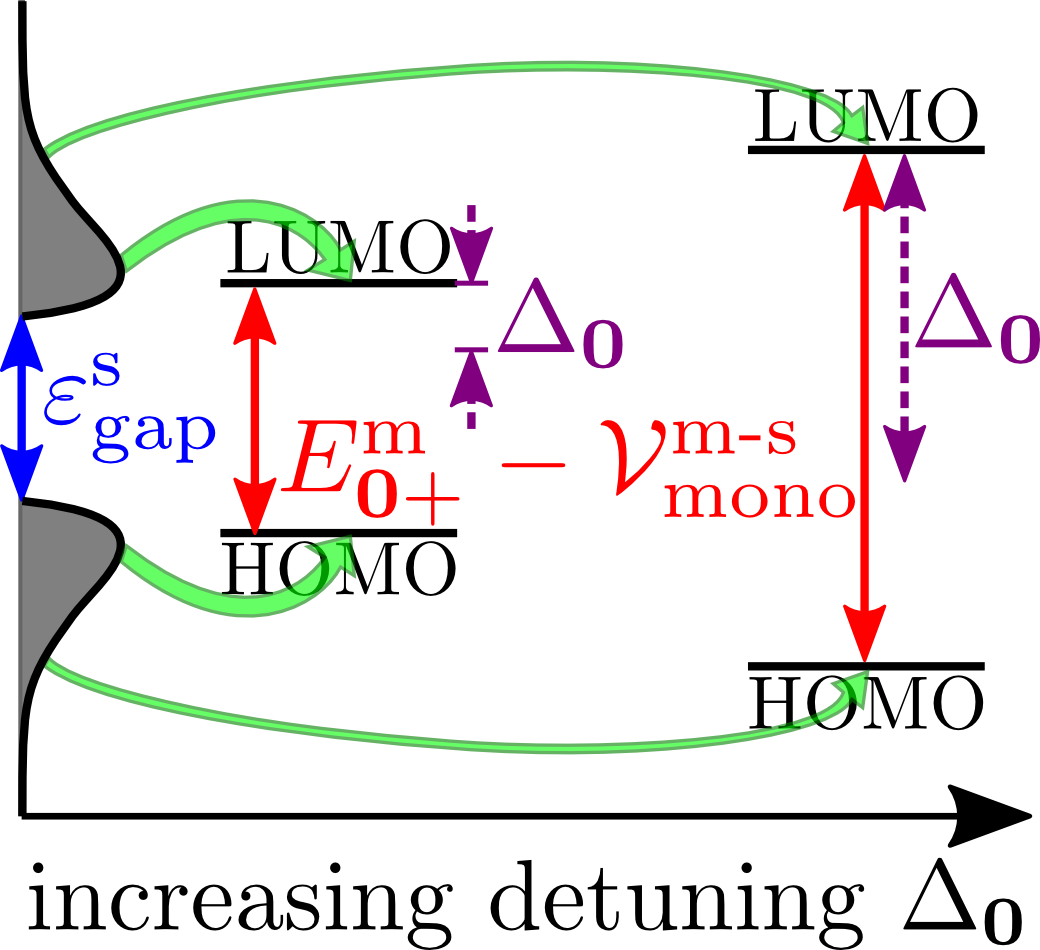}
		(a)
	\end{minipage}
	\hfill
	\begin{minipage}[b]{0.275\linewidth}
		\centering
		\includegraphics[width=\linewidth]{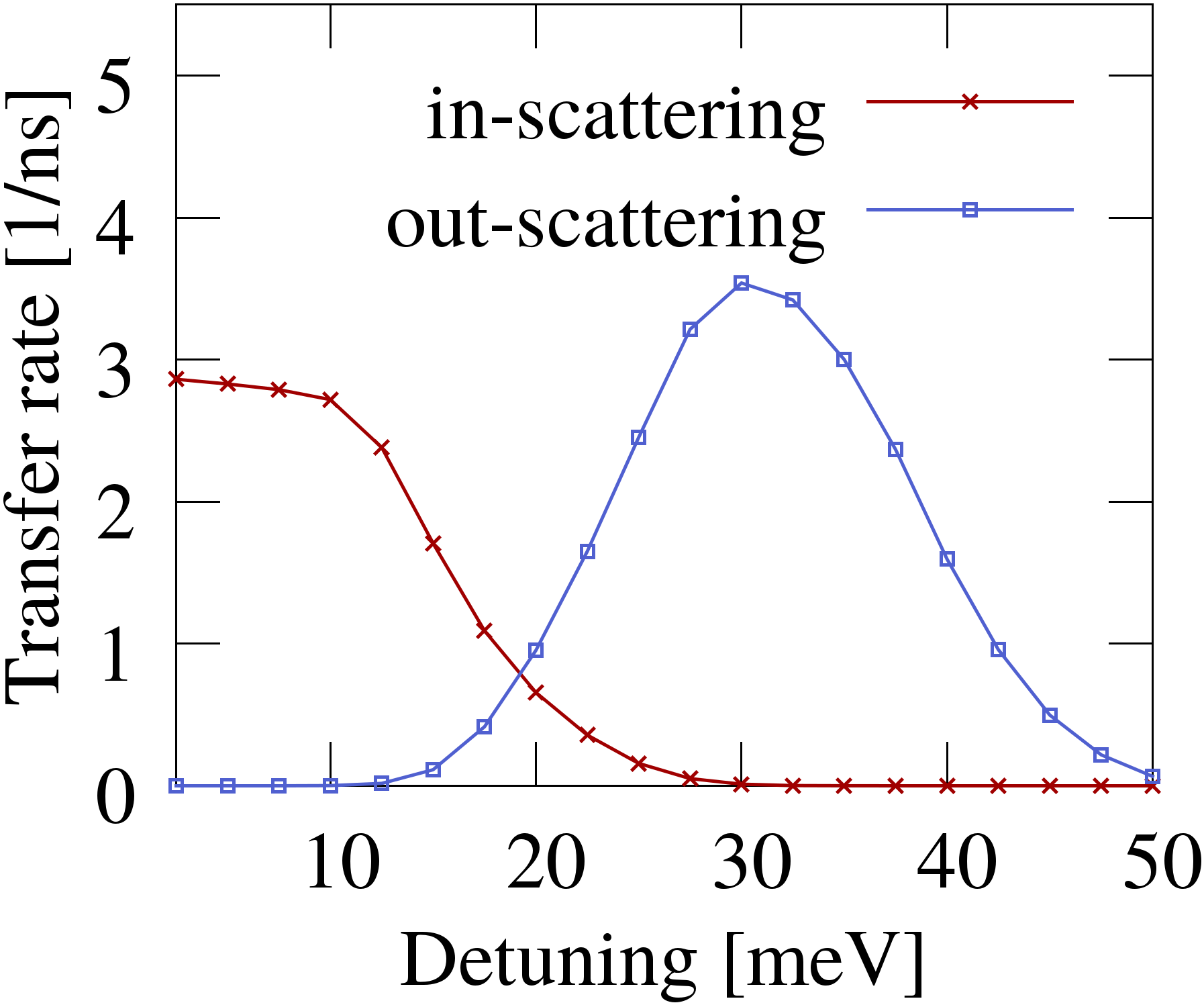}
		(b)
	\end{minipage}
	\hfill\hfill
	\caption{(a) Scheme of the system states of the semiconductor (left) and molecule (right) for increasing detuning $\Delta_{\vec{0}} = E_{\vec{0}}^{\text{m}+} - \varepsilon_\text{gap}^\text{s} - \mathcal{V}_\text{mono}^\text{m-s}$ (not true to scale). (b) Total in- and out-scattering rates from inorganic to organic component as functions of the detuning between the renormalized resonances of the organic and inorganic constituent.}
	\label{fig:TransferRate_Detuning}
\end{figure}

First, we examine the transfer-rate dependence on the detuning $\Delta_{\vec{q}_j = \vec{0}} \equiv \Delta_{\vec{0}} = E_{\vec{q}_j=\vec{0}}^{\text{m}+} - \varepsilon_\text{gap}^\text{s} - \mathcal{V}_\text{mono}^\text{m-s}$ between the renormalized resonances of the two constituents (cf. Fig.~\ref{fig:TransferRate_Detuning} (a)). It enters the energy conserving delta distribution in Eqs.~\eqref{eq:TransferRate} and \eqref{eq:TransferRate_bs}.
Note that the intermolecular monopole-monopole coupling leads to a substantial energy renormalization for high molecular coverages. In the case of the closest molecular packing without steric overlap of one molecule per $6 \times 2$ substrate unit cells (cf. Fig.~\ref{fig:UCs_BZs_6x2}), we adjust the molecular gap $\varepsilon_\text{gap}^\text{m}$ by several tens of $\unit{meV}$ in order to get the two exciton systems into resonance. Of course, in the case of sparse molecular coverages, the intermolecular monopole-monopole shifts have a much weaker effect. The inorganic-organic resonance energy detuning can be controlled by tuning the molecular structure (e.g., exchanging ligands etc. \cite{Schlesinger:NatCommun:15}). Furthermore, other corrections may be important here beyond our simple model which only focuses on the transfer. Therefore, a variation of $\Delta_{\vec{0}}$ (cf. Fig.~\ref{fig:TransferRate_Detuning} (a)) is justified for obtaining a qualitative understanding of the involved transfer processes in theses situations.

Figure~\ref{fig:TransferRate_Detuning} (b) shows the total in- and out-scattering rates as function of the detuning $\Delta_{\vec{0}}$. The rates are in the range of several $\unit{ns^{-1}}$. These values are consistent with experimentally measured transfer times of 100 to $\unit[300]{ps}$ in similar hybrid structures \cite{Blumstengel:PhysRevLett:06,Schlesinger:NatCommun:15}.
The in-scattering rate into the molecular film decreases for increasing $\Delta_{\vec{0}}$ and vanishes for detunings larger than $\unit[30]{meV}$, so that a device operation up to $10$-$\unit[15]{meV}$ should be efficient. In contrast, the out-scattering rate has a maximum around $\unit[30]{meV}$ and drops to zero towards higher ($\sim \unit[50]{meV}$) and lower detunings ($\sim \unit[10]{meV}$).
This behavior can be understood using the scheme of system states shown in Fig.~\ref{fig:TransferRate_Detuning} (a): The carrier population in the semiconductor is depicted along the $x$ axis as product of the Fermi function and the density of states. The molecular states are discrete HOMO and LUMO levels with two different detunings $\Delta_{\vec{0}}$. (This ignores that also the molecular system exhibits a flat band structure due to the intermolecular Coulomb coupling. However, the molecular bands cover a very small energetic range compared to the electrically pumped semiconductor states.)
Low detunings mean a close energetic match between the resonances of the two constituents. This enables an efficient in-scattering into the molecular layer, since the semiconductor substrate exhibits a high population filling where energy and momentum conservation are fulfilled. For higher detunings and increased energy mismatch (cf. right-hand side of Fig.~\ref{fig:TransferRate_Detuning} (a)), the number of available scattering partners in high-energy band states decreases. Therefore, the in-scattering rate shows a strong decrease.
Up to $\Delta_{\vec{0}} = \unit[30]{meV}$, the out-scattering rate in Fig.~\ref{fig:TransferRate_Detuning} (b) shows the opposite behavior dictated by the Pauli blocking terms that prevent back-scattering. However, Pauli blocking gets weaker with increased detuning and the out-scattering rate increases until the energy mismatch between the molecular and semiconductor gap is too large to be bridged by any of the populated states in the semiconductor electron-hole continuum.

\subsection{Influence of the molecular coverage} \label{sec:Coverage}

\begin{figure}
	\centering
	\includegraphics[width=0.45\linewidth]{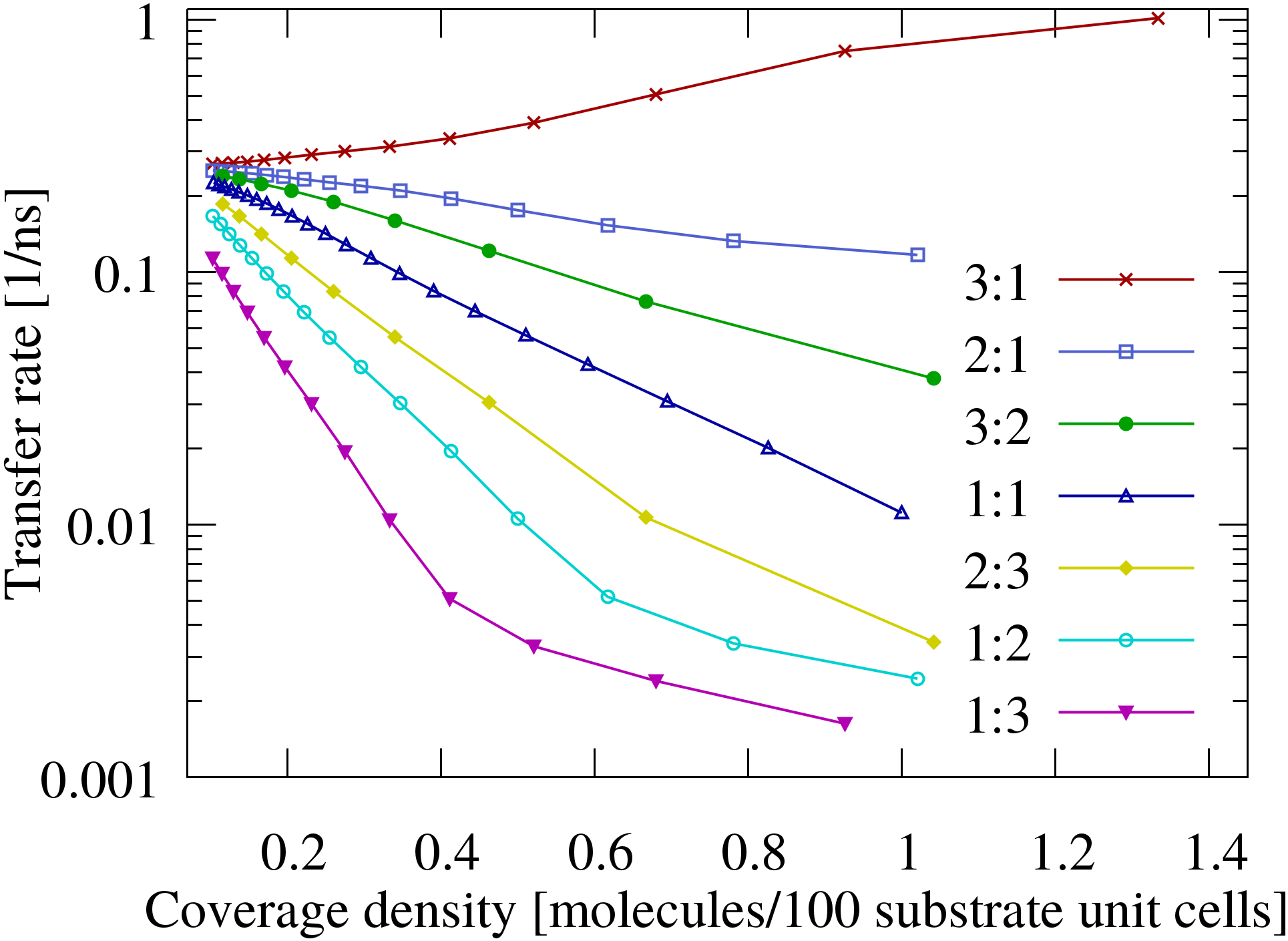}
	\caption{In-scattering rates as functions of the molecular coverage density for different aspect ratios of the molecular coverage.}
	\label{fig:TransferRate_CoverageDensity}
\end{figure}

Figure~\ref{fig:TransferRate_CoverageDensity} shows the excitation energy transfer rate from the electrically pumped semiconductor substrate into the molecular layer as a function of the molecular coverage density for different aspect ratios of molecular coverage. The aspect ratio $n_x:n_y$ defines the ratio between the number of semiconductor unit cells matching a molecular unit cell in $x$ direction and the number in $y$ direction. We calculate the transfer rate for a detuning of $\Delta_{\vec{0}} = \unit[30]{meV}$, since here a calculation over a large parameter range is feasible. Other detunings show the same overall qualitative behavior.
For decreasing molecular coverages, the size of the molecular unit cell in real space increases, whereas the molecular Brillouin zone decreases \cite{Verdenhalven:PhysRevB:14}: The molecular reciprocal grid points get denser until a quasi-continuuous density of reciprocal lattice vectors is achieved. This increases the interlayer Coulomb coupling per molecule in the case of low molecular coverages, since more processes fulfill momentum conservation. For very small coverages (left-hand side of Fig.~\ref{fig:TransferRate_CoverageDensity}), the distance between two neighboring molecules is so large that they do not interact and the particular unit cell geometry defined by the aspect ratio is irrelevant. Therefore, the transfer rates tend towards the same low-coverage limit independent of the aspect ratio.
However, when going towards higher molecular coverages, the transfer efficiency strongly depends on the aspect ratio. Here, two processes are counteracting: On the one hand, increasing the total number of molecules per $100$ substrate unit cells improves the coupling to the substrate, since the coverage density given as the number of molecules divided by the number of semiconductor unit cells $N_\text{m}/N_\text{uc}$ enters the rate. On the other hand, the number of allowed momentum transfer processes decreases for higher coverages due to the decreasing reciprocal grid density.
The dependence on the aspect ratio reflects the spatial orientation of the molecular transition dipole moment $\vec{d}_{\Ho \Lu}$ along the $x$ axis (cf. Fig.~\ref{fig:FoersterCouplingElement} (a)).
For aspect ratios less than $1$, the unit cell has a larger extent in the $y$ direction than in the $x$ direction, thus increasing the number of unfavorable scattering channels perpendicular to the dipole moment. This leads to a weaker interaction for smaller aspect ratios. Indeed, the transfer rate decreases by orders of magnitude when increasing the molecular coverage inside typical ranges (normalized to the molecule number). This is of course negative for the device performance. The main reason for the decrease is the smaller number of allowed transfer processes.
For aspect ratios greater than $1$ that coincide with the dipole orientation, this effect is strongly attenuated or even inverted in the case of a $3:1$ coverage. Here, the increase in transfer efficiency for high coverages dominates over the counteracting decrease of allowed momentum transfer processes.
Note that the maximum molecular coverage without steric overlap is one molecule per $6 \times 2$ substrate unit cells with an aspect ratio of $3:1$ (cf. red curve in Fig.~\ref{fig:TransferRate_CoverageDensity}). This configuration will be most likely in the experiment with one or even multiple fully closed organic layers on top. Here we could show that a dense coverage combined with a high aspect ratio as in the case of maximum coverage is advantageous for the device performance.

\subsection{Tuning the electrical driving: influence of the carrier concentration} \label{sec:CarrierConcentration}

\begin{figure}
	\centering
	\includegraphics[width=0.35\linewidth]{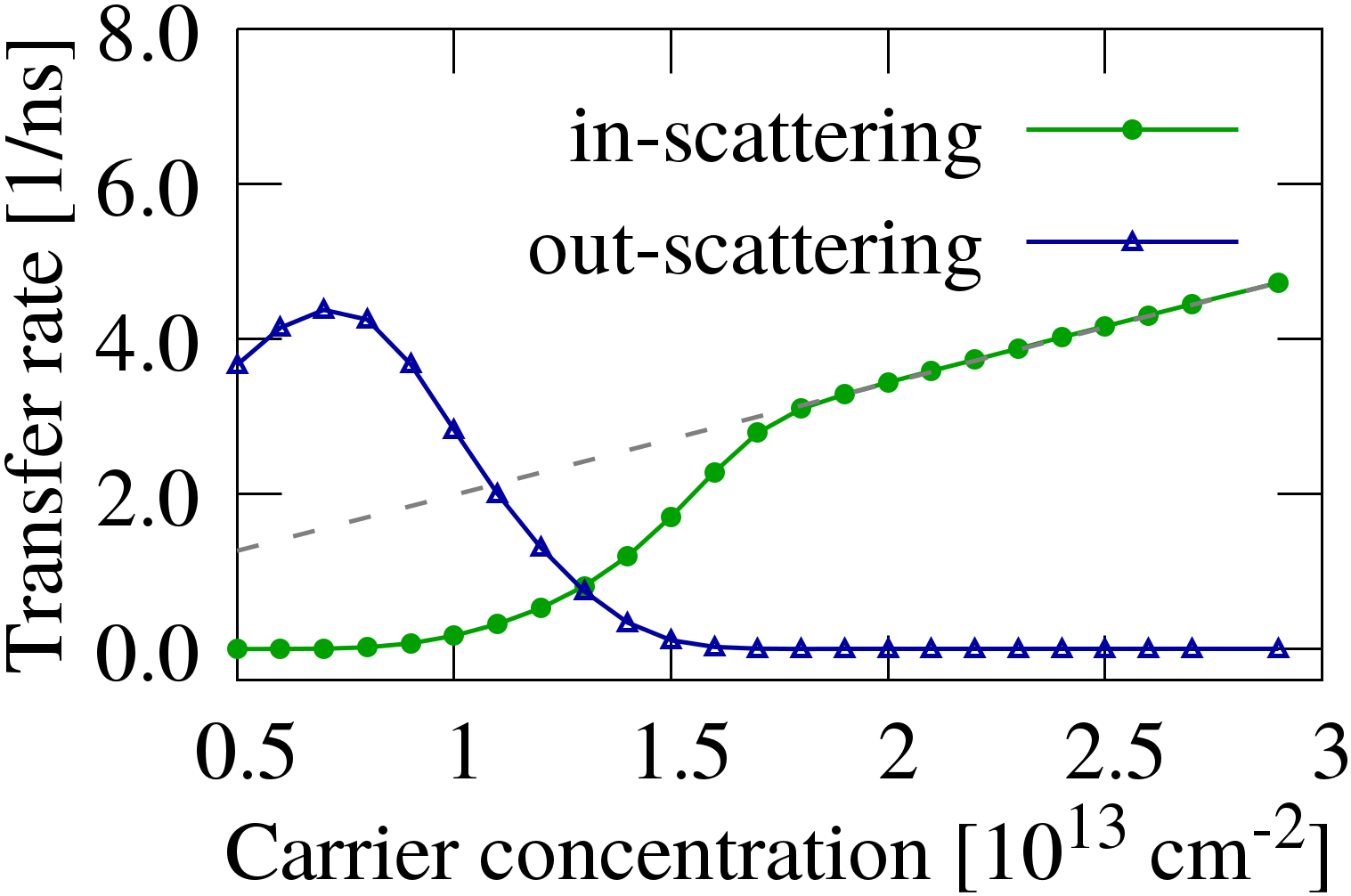}
	\caption{Total in- and out-scattering rates as functions of the two-dimensional carrier concentration in the semiconductor QW for an aspect ratio of $1:1$ ($10 \times 10$ coverage) and charge carrier temperature of $\unit[10]{K}$. The dashed gray line marks the linear regime the in-scattering rate enters at $n_\text{e/h}^\text{2D} = \unit[1.8 \times 10^{13}]{cm^{-2}}$.}
	\label{fig:TransferRate_CarrierConcentration}
\end{figure}

In Fig.~\ref{fig:TransferRate_CarrierConcentration}, the in- and out-scattering transfer rates are depicted for increasing carrier concentrations in the semiconductor part. Through the chemical potential in Eq.~\eqref{eq:chemicalPotential}, the carrier densities enter the Fermi distribution of the electron and hole continuum in the inorganic constituent. The transfer efficiency is highly sensitive to the charge carrier concentration (and thereby, the electrical pump strength): As expected, the in-scattering rate per molecule becomes larger for increasing carrier densities due to the higher number of carriers that are available as scattering partners. This increases the number of carriers with energies fulfilling energy conservation for transfer and thus provides an increased transfer efficiency. First, the transfer rate increases non-linearly, then it enters a period of linear growth at $n_\text{e/h}^\text{2D} = \unit[1.8 \times 10^{13}]{cm^{-2}}$ indicated by the dashed gray line in Fig.~\ref{fig:TransferRate_CarrierConcentration}. The initial non-linear growth of the rate for low carrier concentrations is attributed to the energy and momentum conservation: At low $n_\text{e/h}^\text{2D}$, only few scattering channels are available in the absence of higher energy and momentum states. With increasing carrier concentrations, the number of possible scattering partners increases until the momentum and energy allowed interaction channels are saturated. Then, the rate enters the linear growing regime dictated simply by the constant growth of the carrier density.
Surprisingly, up to $n_\text{e/h}^\text{2D} = \unit[0.7 \times 10^{13}]{cm^{-2}}$, also the out-scattering rate increases before decreasing again. The unexpected initial growth of the out-scattering rate is explained as follows: For very low carrier densities, only electronic states close to the $\Gamma$ point are populated, whereas higher energy and momentum states are not occupied, cf. Fig.~\ref{fig:Fermiplot} (c) and (d). This restricts the possible transfer processes to a small energy and momentum range, thus reducing both the in- and out-scattering excitation transfer efficiency. For increasing carrier concentrations, more electronic states contribute. However, at a certain carrier concentration, Pauli blocking is reached in the semiconductor QW and reduces the out-scattering rate again: For larger carrier concentrations, an increasing number of electron-hole continuum states is occupied, thus preventing back-scattering into the semiconductor layer, i.e., this process becomes negligibly small.

\subsection{Tuning the electrical driving: changing the carrier temperature} \label{sec:CarrierTemperature}

\begin{figure}
	\centering
	\includegraphics[width=0.495\linewidth]{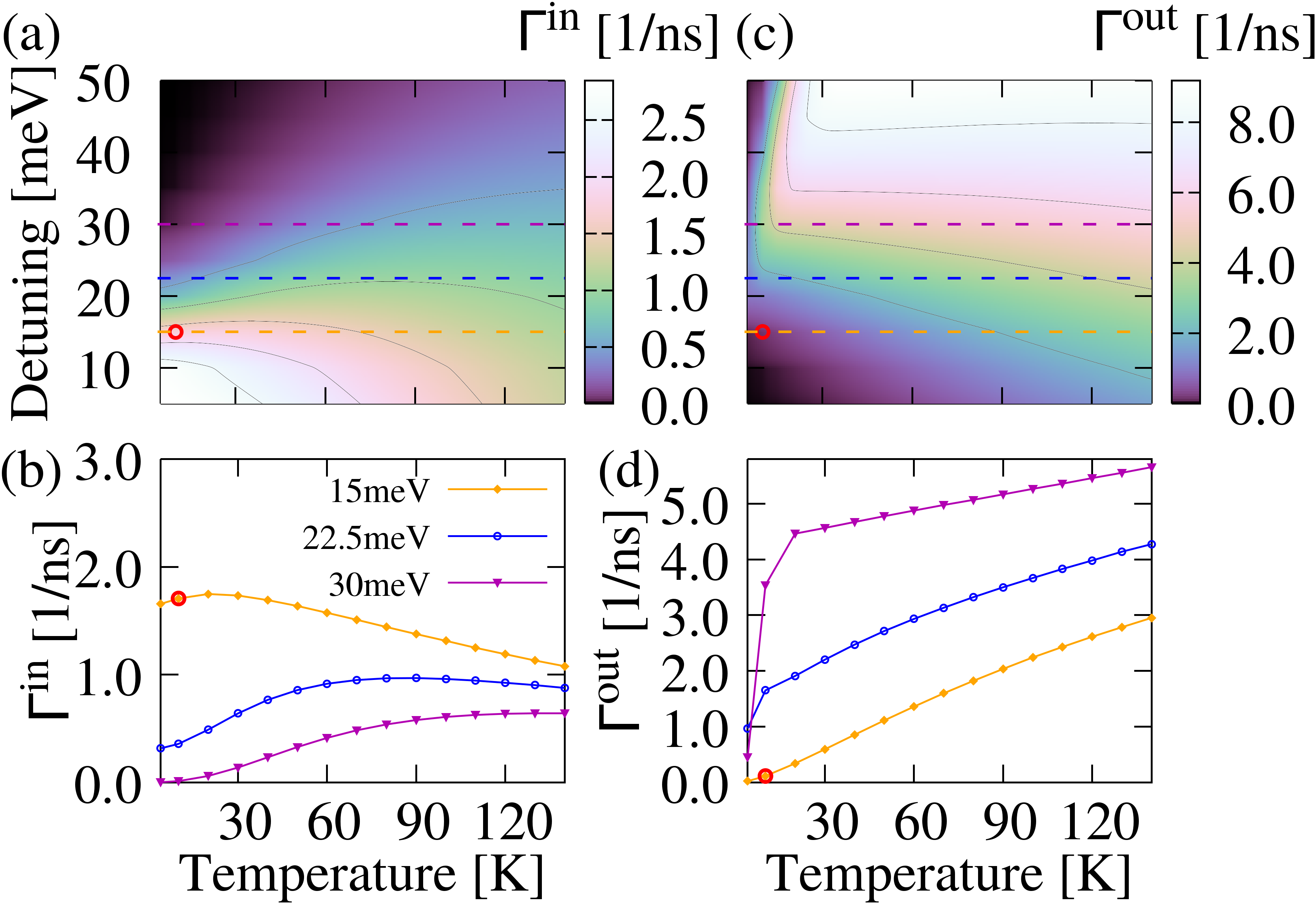}
	\caption{Total in-scattering (a) and out-scattering rate (c) as a function of the charge carrier temperatur $T = T_\text{e} = T_\text{h}$ and energetic detuning $\Delta_{\vec{0}}$. The lower panels (b) and (d) show cuts through the 2D plots (a) and (c) (indicated by the dashed gray lines) at fixed detunings $\Delta_{\vec{0}} = \unit[15]{meV}$ (orange curves), $\unit[22.5]{meV}$ (blue curves), and $\unit[30]{meV}$ (purple curves). The red circles mark the standard values used throughout this work (cf. Tab. \ref{tab:parameters}).}
	\label{fig:TransferRates_Temperature}
\end{figure}

Figure~\ref{fig:Fermiplot} (a) and (b) show that the temperature of the charge carriers in the QW changes the carrier distribution in the semiconductor bands considerably. To analyze the interplay between the resonance energy detuning $\Delta_{\vec{0}}$ and the temperatures $T_\text{e/h}$, we calculated 2D maps for the in-scattering (Fig.~\ref{fig:TransferRates_Temperature} (a)) and out-scattering rates (Fig.~\ref{fig:TransferRates_Temperature} (c)) in dependence of temperature and detuning. For higher temperatures, the in- and outgoing rates are less sensitive to resonance energy detunings and the transfer efficiency is less dependent on the temperature. Here, the increased population of high energy band states in the QW at higher temperatures results in increased energy matching. This leads to a monotonous increase of the back-scattering rate with increasing temperature, cf. Fig.~\ref{fig:TransferRates_Temperature} (c) and (d).
However, an interesting feature occurs in the case of the in-scattering rate, cf. Fig.~\ref{fig:TransferRates_Temperature} (a). To highlight this effect, we additionally plotted cuts through the 2D map at fixed detunings $\Delta_{\vec{0}} = \unit[15]{meV}$ (orange curve), $\unit[22.5]{meV}$ (blue curve), and $\unit[30]{meV}$ (purple curve), cf. Fig.~\ref{fig:TransferRates_Temperature} (b). The positions of the cuts are marked by the gray dashed lines in the 2D map in Fig.~\ref{fig:TransferRates_Temperature} (a). For larger detunings $\Delta_{\vec{0}} > \unit[25]{meV}$ (upper region in the 2D map of Fig.~\ref{fig:TransferRates_Temperature} (a) and purple curve in the graph of Fig.~\ref{fig:TransferRates_Temperature} (b)), the in-scattering rate increases monotonously with increasing temperature, as one would expect due to the population of higher electronic states fulfilling the energy conservation condition. In contrast, for lower detunings $\Delta_{\vec{0}} < \unit[25]{meV}$, the transfer rate shows a slight increase at first but then drops down again when going towards higher temperatures. Small detunings between the semiconductor band gap and molecular gap require a close energetic match that is only fulfilled by lower energy states close to the $\Gamma$ point. However, in the high-temperature regime, the population of the electron-hole continuum states close to the $\Gamma$ point at $\vec{k} = 0$ decreases, whereas in turn higher energy and momentum states are occupied (cf. Fig.~\ref{fig:Fermiplot}). This decreases the transfer efficiency in the case of high temperatures and small detunings.
The back-scattering rate in Fig.~\ref{fig:TransferRates_Temperature} (b) shows the opposite behavior: The higher the detuning, the higher the back-scattering efficiency. However, this only holds for temperatures above $\unit[30]{K}$. Below that, the out-scattering decreases for higher detunings, see also the non-monotonous shape of the out-scattering curve in Fig.~\ref{fig:TransferRate_Detuning} (b) for $T = \unit[10]{K}$. Here, the large energetic detuning between the inorganic and organic part counteracts the fact that at low temperatures only low energy band states are populated.

\subsection{Variation of the orientation and distance of the molecular film} \label{sec:OrientationDistance}

\begin{figure}
	\centering
	\includegraphics[width=0.48\linewidth]{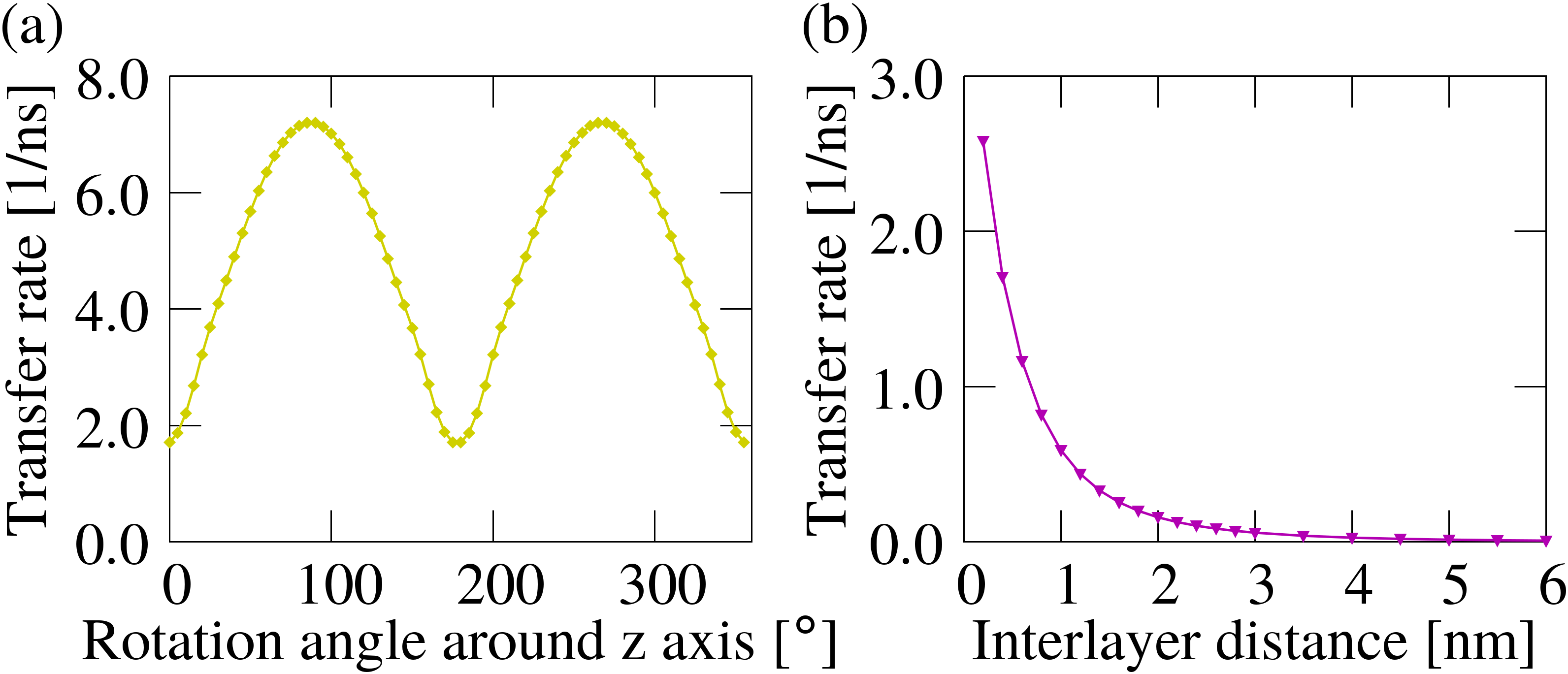}
	\caption{(a) Total in-scattering rate as a function of the rotational angle of the molecules around the $z$ axis. (b) Total in-scattering rates as functions of the interlayer separation between the semiconductor QW and the molecular layer.}
	\label{fig:TransferRate_Rot_Dist}
\end{figure}

As discussed in Sec.~\ref{sec:PCA}, the validity of the partial charge technique exceeds the dipole-dipole approximation. However, effective transition dipole moments can be assigned to the partial charges belonging to one constituent according to Eqs.~\eqref{eq:totalDipoleMomentS} and \eqref{eq:totalDipoleMomentM}: $\vec{d}_{\cb \vb} = \sum_i q_i^{\cb \vb}\vec{r}_i$ and $\vec{d}_{\Lu \Ho} =\sum_j q_j^{\Lu \Ho}\vec{r}_j$. Therefore, we expect a dependence of the transfer efficiency on the orientation of the molecules on top of the semiconductor surface. In the geometry found by DFT calculations, the effective dipole moments of the organic and inorganic constituent lie both in the QW plane, however, they are oriented almost perpendicular. Therefore, we rotate the molecules within their plane around the vertical $z$ axis. Figure~\ref{fig:TransferRate_Rot_Dist} (a) depicts the transfer rate into the organic film in dependence of the rotational angle of the molecules around the $z$ axis. Indeed, we observe a $\cos^ 2$-like behavior with maxima at roughly $85 \degree$ and $265 \degree$, where the effective dipole moments are approximately parallel. This reflects the interlayer F\"orster coupling element entering the rate in Eq.~\eqref{eq:TransferRate} squared. Note that also other parameters besides the simple $\cos$ dependence of the interlayer coupling element play a role when rotating the molecules, since also the molecular band dispersion and therefore the energy matching condition is altered due to the changed intermolecular coupling. Also, the rate does not drop to zero for perpendicular effective dipole moments as one would expect for a pure dipole interaction, since there is always a substantial remaining coupling strength due to the spatial distribution of the partial charges. At perpendicular dipole moments, the rate still is around $24 \%$ of the maximum value at parallel dipoles.

Figure~\ref{fig:TransferRate_Rot_Dist} (b) shows the transfer rate into the molecular layer for increasing distance $\Delta z$ between the semiconductor substrate and the molecular adlayer. Different separations between the QW and the adsorbed organic layer can be realized experimentally, e.g., by inserting a spacer layer of variable thickness \cite{Blumstengel:PhysRevLett:06}. As expected, we observe a strong decrease of the transfer efficiency with increasing interlayer separation, since the interlayer F\"orster coupling strength entering the transfer rate is decreasing for increasing distances. Note that we did not depict the back-scattering rate: it shows the same qualitative behavior.

\section{Conclusions} \label{chapter:Conclusions}

In this manuscript, we used a microscopic theory for calculating the energy transfer rate from an electrically pumped inorganic semiconductor substrate into an organic molecular film. Partial charges for both the semiconductor and the molecules were obtained from DFT calculations of the electrostatic potential and were used to model the microscopic coupling elements beyond the common dipole-dipole treatment. We showed that the transfer efficiency can be improved substantially by altering the geometric arrangement of the hybrid system and by varying the pumping strength. The effect of F\"orster coupling between the two layers is governed by the orientation of the dipole moments in the two constituents and microscopic momentum and energy selection rules, making HIOS highly versatile building blocks for device application.

For optimizing future devices, we recommend the following guidelines: (i) Near-field effects should be exploited by using short distances and the dipoles of molecules and semiconductor should be aligned. (ii) In order to suppress the back-scattering, further layers of molecules with smaller band gap should be added to act as a cascade. (iii) The carrier concentration operating point should be high enough that in-scattering outweighs out-scattering and the transition energy of molecules and semiconductor should be aligned accordingly.

\section*{Acknowledgment}

We gratefully acknowledge financial support from the Deutsche Forschungsgemeinschaft (DFG) through SFB 951. This work was supported by the Academy of Finland through its Centres of Excellence Programme (2012-2014 and 2015-2017) under project numbers 251748 and 284621.

\begin{appendices}

\section{Construction of partial charges} \label{App1:PartialCharges}

The parameterization of the density matrix formalism from the electronic structure obtained by DFT calulations has previously been developed for purely crystalline semiconductor surfaces \cite{Buecking:PhysRevB:08}. The optical excitations and electron relaxation dynamics of silicon surfaces were calculated by a DFT parameterization based on the band structure, momentum-matrix elements and phonon band structure obtain\-ed with a (semi-)local xc-functional. The first order approximation for the Coulomb-matrix elements is equivalent to the dipole-matrix elements between the semiconductor and the molecular system.

The challenge is to find reliable partial charges for molecule and semiconductor.  A simple method for cluster calculations (molecules) as well as two methods for solids (periodic boundary conditions) \cite{Campana:JChemTheoryComput:09, Chen:JPhysChemA:10} were implemented in the FHI-aims code \cite{Blum:ComputPhysCommun:09}.

The starting point for these methods is the calculation of the electrostatic potential at a sufficiently high number of grid points outside the van der Waals (vdW) radius of the atoms \cite{Pauling::60} (defined as the radius of imaginary hard spheres reflecting the contact distance of the atoms \cite{Pauling::60}). To define a spatial region for the grid, two parameters are necessary: a minimal and a maximal radius around the atoms. These radii are defined as multiples of the vdW-radius of the atoms. The values for the vdW radii of most atoms in the periodic table have been taken from Refs.~\onlinecite{Bondi:JPhysChem:64, Rowland:JPhysChem:96, Mantina:JPhysChemA:09}. For the generation of the points cubic (Cartesian) grids are used. For finite systems such as molecule, we generate points within a cube encapsulating the spheres with the maximal radius (multiple of the vdW radius) around all atoms. For periodic boundary conditions, the unit cell is used. Points within the superposition of the spheres with the minimal radius (minimal multiple of the vdW radius) are excluded.

For finite systems we express the electrostatic potential (ESP) by a sum of Coulomb potentials with charges $q_i$, the partial charges, at the atomic position $\vec{r}_i$:
\begin{equation}
	 V_\text{ESP}^\text{mol} (\vec{r})=\sum_{i=1}^{N_\text{at}}\frac{q_i}{\left\lvert \vec{r}-\vec{r}_i \right\rvert}.
\end{equation}
The $q_i$ are calculated by a least squares fit with the additional constraint of constant total charge $q_\text{tot}=\sum_{i=1}^{N_\text{at}} q_i$. 
We use the method of Lagrange multipliers to minimize the function
\begin{equation} \label{F_esp}
	F=\sum_{k=1}^{N_\text{grid}}\left(V_\text{ES,DFT}^\text{mol} (\vec{R}_k)-V_\text{ESP}^\text{mol}(\vec{R}_k)\right)^2-\lambda\left(q_\text{tot}-\sum_{i=1}^{N_\text{at}}q_i\right)^2
\end{equation}
where $V_\text{ES,DFT}$ is the electrostatic potential of a DFT calculation.

For periodic systems, long-range electrostatic interactions have to be taken into account and the potential is only defined up to an arbitrary constant  \cite{Campana:JChemTheoryComput:09}. The partial charge methods implemented in this work solve these problem by Ewald summation \cite{Ewald:AnnalenderPhysik:21}. They were developed by Campana \textrm{et~al.} \cite{Campana:JChemTheoryComput:09} and further improved by Chen \textrm{et~al.} \cite{Chen:JPhysChemA:10}. The potential generated by the partial charges centered on the atoms of the unit cell then reads
\begin{align}
	\begin{split} \label{V_Ewald}
		V_\text{ESP}^\text{solid} (\vec{r})=&\sum_{i=1}^{N_\text{at}} \sum_{\vec{T}} q_i\frac{\mathrm{erfc}(\alpha\left\lvert\vec{r}-\vec{r}_{i,\vec{T}}\right\rvert)}{\left\lvert\vec{r}-\vec{r}_{i,\vec{T}}\right\rvert}\\
		&+\frac{4\pi}{V_\text{uc}}\sum_{i=1}^{N_\text{at}} \sum_{\vec{k}} q_i\mathrm{cos}(\vec{k}(\vec{r}-\vec{r}_{i}))\frac{\mathrm{e}^{-\frac{k^2}{4\alpha^2}}}{k^2}
	\end{split}
\end{align}
where $\vec{T}=n_1\vec{a}_1+n_2\vec{a}_2+n_3\vec{a}_3$ are real space translation vectors of the lattice vectors $\vec{a}_i$ and $n_i \in  \mathbb{Z}$. $\vec{k}=m_1\vec{b}_1+m_2\vec{b}_2+m_3\vec{b}_3$ is the reciprocal space translation vector and $\vec{b_i}$ are the reciprocal lattice vectors with $m_i \in \mathbb{Z}$. $V_\text{uc}$ is the volume of the unit cell. The parameter $\alpha$ is defined as $\alpha=\frac{\sqrt{\pi}}{R_\text{c}}$ with $R_\text{c}$ the cutoff radius of the Ewald summation \cite{Ewald:AnnalenderPhysik:21}. The function to minimize is \cite{Chen:JPhysChemA:10}:
\begin{align}
	\begin{split} \label{F_esp_pbc2}
		F^\text{PBC}=&\sum_{k=1}^{N_\text{grid}}\left(V_\text{DFT}^\text{solid}(\vec{R}_k)-\left(V_\text{ESP}^\text{solid}(\vec{R}_k)+V_\text{offset}\right)\right)^2\\
		&-\lambda\left(q_\text{tot}-\sum_{i=1}^{N_\text{at}}q_i\right)+\beta\sum_{i=1}^{N_\text{at}}\left(q_i-q_{i0}\right)^2.
	\end{split}
\end{align}
Here the arbitrary offset of the potential $V_\text{offset}$ is an additional fitting parameter. The constraint charges $q_{i0}$ can be determined with other methods (e.~g. Mulliken charge analysis \cite{Mulliken:JChemPhys:55}). $\beta$ is a weighting factor.

\section{Calculation of the eigenproblem of the molecular excitons} \label{App2:MolecularEigenproblem}

The eigenproblem for the molecular excitons reads
\begin{equation}
	\bra{\vec{l} + \vec{q}, \vec{l}} \hat{H}^\text{m}_0 + \hat{H}^\text{m-m}_\text{C} \ket{X_\alpha^\text{m}}
	= (E_0^\text{m} + E_\alpha^\text{m}) \; c^\alpha_{\vec{l} + \vec{q}, \vec{l}}.
\end{equation}

For a sufficiently large material sample, the molecular wave vectors are continuous and the sums can be transformed into two-dimensional integrals according to
\begin{equation}
	\sum_{\vec{l}} \rightarrow \frac{N_\text{m} A_\text{m}}{(2 \pi)^2}\int \text{d}^2 l,
	\label{eq:42}
\end{equation}
where $A_\text{m}$ denotes the area of one molecular unit cell.
Exploiting the lattice periodicity of the Coulomb coupling elements in momentum space, we find:
\begin{align}
	\begin{split} \label{eq:molEigenproblem2}
		E_\alpha^\text{m} & c^\alpha_{\vec{l}+\vec{q}, \vec{l}} = 
		c^\alpha_{\vec{l} + \vec{q}, \vec{l}} \Bigl[ \varepsilon_\text{gap}^\text{m} - \mathcalV{\Ho}{\Ho}{\Ho}{\Ho} (\vec{0}) + \mathcalV{\Ho}{\Lu}{\Ho}{\Lu} (\vec{0}) \\
		+ & \frac{A_\text{m}}{4 \pi^2} \int \text{d}^2 l^\prime \; \Bigl( \mathcalV{\Ho}{\Ho}{\Ho}{\Ho} (\vec{l}^\prime) - \mathcalV{\Ho}{\Lu}{\Lu}{\Ho} (\vec{l}^\prime) \Bigr) \Bigr] \\
		+ & \frac{A_\text{m}}{4 \pi^2} \int \text{d}^2 l^\prime \; c^\alpha_{\vec{l}^\prime + \vec{q}, \vec{l}^\prime} \Bigl[  \mathcalV{\Ho}{\Lu}{\Lu}{\Ho} (\vec{q}) - \mathcalV{\Ho}{\Lu}{\Ho}{\Lu} (\vec{l}^\prime - \vec{l}) \Bigr]
	\end{split}
\end{align}
with $\varepsilon_\text{gap}^\text{m} = \varepsilon_\Lu - \varepsilon_\Ho$.
To make the problem numerically tractable, the continuous wave vectors are discretized. Therefore, the integrals over the first BZ are rewritten into sums over $N_\text{d}^\text{m}$ small surface segments of size $\Delta A \equiv \frac{A_\text{BZ}^\text{m}}{N_\text{d}^\text{m}}$:
\begin{equation}
	\int \text{d}^2 l^\prime \; f(\vec{l}^\prime) \rightarrow \sum_{i = 1}^{N_\text{d}^\text{m}} \Delta A f (\vec{l}_i).
\end{equation}
Moreover, we approximate $\mathcalV{\Ho}{\Lu}{\Ho}{\Lu} (\vec{l}^\prime - \vec{l}) \approx \mathcalV{\Ho}{\Lu}{\Ho}{\Lu} (\vec{0})$ in the last line of Eq.~\eqref{eq:molEigenproblem2}, since the variation of the monopole-monopole coupling elements $\mathcalV{\Ho}{\Lu}{\Ho}{\Lu}$ within the first BZ is small (only few percent). This reduces the complexity of the eigenproblem and we can derive the eigenproblem for the energy $E_\alpha^\text{m}$ and the coefficients $c^{\vec{q}}_{\vec{l}_i + \vec{q}_j, \vec{l}_i}$ in matrix form. It is diagonal with respect to the momentum transfer $\vec{q}_j$, yielding a block-diagonal form for the entire index space $(\vec{l}_i, \vec{q}_j)$.
We use the abbreviations:
\begin{align}
	a_{\vec{q}_j} = & \frac{1}{N_\text{d}^\text{m}} \left[ \mathcalV{\Ho}{\Lu}{\Lu}{\Ho} (\vec{q}_j) - \mathcalV{\Ho}{\Lu}{\Ho}{\Lu} (\vec{0}) \right], \label{eq:offdiagonalElements} \\
	\begin{split} \label{eq:diagonalElements}
		d_{\vec{q}_j} \equiv & a_{\vec{q}_j} + \varepsilon_\text{gap}^\text{m} - \mathcalV{\Ho}{\Ho}{\Ho}{\Ho} (\vec{0}) + \mathcalV{\Ho}{\Lu}{\Ho}{\Lu} (\vec{0}) \\
		& + \frac{1}{N_\text{d}^\text{m}} \sum_{k=1}^{N_\text{d}^\text{m}} \; \left[ \mathcalV{\Ho}{\Ho}{\Ho}{\Ho} (\vec{l}_k) - \mathcalV{\Ho}{\Lu}{\Lu}{\Ho} (\vec{l}_k) \right]
	\end{split}
\end{align}
Each $N_\text{d}^\text{m} \times N_\text{d}^\text{m}$ block for a given $\vec{q}_j$ has the form:
\begin{align}
	\begin{split}
		E_{\vec{q}_j}^\text{m}
		\begin{pmatrix}
			c_{\vec{l}_1}^{\vec{q}_j} \\
			c_{\vec{l}_2}^{\vec{q}_j} \\
			\vdots \\
			c_{\vec{l}_{N_\text{d}^\text{m}}}^{\vec{q}_j}
		\end{pmatrix}
		=
		\begin{pmatrix}
			d_{\vec{q}_j} & a_{\vec{q}_j} & \cdots & a_{\vec{q}_j} \\
			a_{\vec{q}_j} & \ddots & & \vdots \\
			\vdots & & \ddots & a_{\vec{q}_j} \\
			a_{\vec{q}_j} & \cdots & a_{\vec{q}_j} & d_{\vec{q}_j}
		\end{pmatrix}
		\begin{pmatrix}
			c_{\vec{l}_1}^{\vec{q}_j} \\
			c_{\vec{l}_2}^{\vec{q}_j} \\
			\vdots \\
			c_{\vec{l}_{N_\text{d}^\text{m}}}^{\vec{q}_j}
		\end{pmatrix}
	\end{split}
\end{align}
with the eigenvector components abbreviated by $c_{\vec{l}_i}^{\vec{q}_j} \equiv c_{\vec{l}_i+\vec{q}_j, \vec{l}_i}^{\vec{q}_j}$.
Note that they are only defined for wave vector sums $\vec{l}_i+\vec{q}_j$ within the first BZ. If $\vec{l}_i + \vec{q}_j$ exceeds the first BZ, it is mapped back into the first BZ by means of a reciprocal lattice vector.

This highly symmetric eigenproblem in matrix form can be solved analytically. It has two eigenvalues:
\begin{align}
	E_-^ \text{m} & = E_{\vec{q}_j, n = 1}^{\text{m}} = \cdots = E_{\vec{q}_j, n = N_\text{d}^\text{m} - 1}^{\text{m}} = d_{\vec{q}_j} - a_{\vec{q}_j}, \\
	E_{\vec{q}_j +}^\text{m} & = E_{\vec{q}_j, n = N_\text{d}^\text{m}}^{\text{m}} = d_{\vec{q}_j} + (N_\text{d}^\text{m} - 1) a_{\vec{q}_j}.
\end{align}
$E_-^\text{m}$ is $(N_\text{d}^\text{m} - 1)$-fold degenerate.

The $(N_\text{d}^\text{m} - 1)$ eigenvectors belonging to $E_{\vec{q}_j}^{\text{m}-}$ are enumerated by $n \in \{1, \dots, N_\text{d}^\text{m} - 1\}$. Their normalized components are given by
\begin{align} \label{eq:antisymmetricCoefficients}
	c_{\vec{l}_i}^{\vec{q}_j, n} = \frac{N_\text{d}^\text{m}}{N_\text{m}} \times
	\begin{cases}
		\frac{1}{\sqrt{2}} \quad \text{for} \; i=1, \\
		- \frac{1}{\sqrt{2}} \quad \text{for} \; i=n+1, \\
		0 \quad \text{else}.
	\end{cases}
\end{align}
The non-degenerate eigenvalue $E_{\vec{q}_j +}^\text{m}$ has identical normalized eigenvector components
$c^{\vec{q}_j, n = N_\text{d}^\text{m}}_{\vec{l}_i} = \tfrac{\sqrt{N_\text{d}^\text{m}}}{N_\text{m}}$ for all $i$.

\end{appendices}


\end{document}